\documentclass[useAMS,usenatbib]{mn2e}
\usepackage{graphicx}
\usepackage{color}

\newcommand{\re}{$\rm R_{\rm e}$}

\newcommand{\ml}{$\rm M/L$}

\newcommand{\simlt}{\lower.5ex\hbox{$\; \buildrel < \over \sim \;$}}
\newcommand{\simgt}{\lower.5ex\hbox{$\; \buildrel > \over \sim \;$}}

\newcommand{\mgf}{$\rm Mg4780$}

\newcommand{\atio}{$\rm aTiO$}
\newcommand{\tioi}{$\rm TiO1$}

\newcommand{\tioii}{$\rm TiO2$}
\newcommand{\tionir}{$\rm TiO0.89$}
\newcommand{\feh}{$\rm FeH0.99$}
\newcommand{\fem}{$\rm \langle Fe\rangle$}

\newcommand{\cfs}{$\rm C4668$}

\newcommand{\mgfep}{$\rm [MgFe]'$}
\newcommand{\mgfe}{$\rm [Mg/Fe]$}
\newcommand{\cafe}{$\rm [Ca/Fe]$}
\newcommand{\nafe}{$\rm [Na/Fe]$}
\newcommand{\tife}{$\rm [Ti/Fe]$}

\newcommand{\cfe}{$\rm [C/Fe]$}
\newcommand{\ofe}{$\rm [O/Fe]$}
\newcommand{\nfe}{$\rm [N/Fe]$}
\newcommand{\sife}{$\rm [Si/Fe]$}
\newcommand{\fet}{$\rm Fe3$}

\newcommand{\hbo}{$\rm H\beta_o$}
\newcommand{\hb}{$\rm H\beta$}
\newcommand{\hgs}{$\rm H\gamma_{275}$}
\newcommand{\mgb}{$\rm Mgb5177$}

\newcommand{\kms}{\,km\,s$^{-1}$}
\newcommand{\afe}{$[\rm \alpha/{\rm Fe}]$}

\newcommand{\mlstar}{$M_\star/L$}
\newcommand{\gammab}{$\rm \Gamma_b$}

\newcommand{\zh}{$\rm [M/H]$}

\newcommand{\dtd}{$\rm \delta(T_{Eff,D})$}
\newcommand{\amiles}{$\rm \alpha \! - \! MILES$}
\newcommand{\datio}{$\rm \delta_{aTiO}$}
\newcommand{\ciz}{$\rm C_{i}(Z_{MW})$}

\title[IMF from TiO indices and Wing-Ford band.]
{Radial constraints on the Initial Mass Function from TiO features and Wing-Ford band in Early-type Galaxies}
\author[La Barbera et al.]
{Francesco  La Barbera$^{1}$\thanks{E-mail: labarber@na.astro.it},
Alexandre Vazdekis$^{2,3}$, Ignacio  Ferreras$^{4}$, Anna Pasquali$^{5}$,\and 
Michele Cappellari$^{6}$, Ignacio Mart\'\i n-Navarro$^{2,3}$,
Frederik Sch\"onebeck$^{5}$, \and
Jes\'us Falc\'on-Barroso$^{2,3}$
\\  
$^{1}$INAF-Osservatorio Astronomico di Capodimonte, sal. Moiariello
16, Napoli, 80131, Italy\\
$^{2}$Instituto de Astrof\'\i sica de Canarias, Calle V\'\i a L\'actea s/n, E-38205
  La Laguna, Tenerife, Spain\\
$^{3}$Departamento de Astrof\'\i sica, Universidad de La Laguna (ULL), E-38206  La Laguna, Tenerife, Spain\\
$^{4}$Mullard Space Science Laboratory, University College London, Holmbury St Mary,
  Dorking, Surrey RH5 6NT, UK\\
$^{5}$Astronomisches Rechen-Institut, Zentrum f\"ur Astronomie, Universit\"at Heidelberg, 
M\"onchhofstr. 12-14, D-69120 Heidelberg, Germany\\
$^{6}$Sub-department of Astrophysics, Department of Physics, University of Oxford, 
Denys Wilkinson Building, Keble Road, Oxford OX1 3RH, UK
}
\begin{document}

\date{Accepted 2015 December 22.  Received 2015 November 22; in original form 2015 September 28}

\pagerange{\pageref{firstpage}--\pageref{lastpage}} \pubyear{2015}

\maketitle

\label{firstpage}

\begin{abstract}
  At present, the main challenge to the interpretation of variations
  in gravity-sensitive line strengths as driven by a non-universal
  initial mass function (IMF), lies in understanding the effect of the
  other population parameters. Most notably, \afe-enhanced populations
  or even departures in the {\sl individual} element abundances with
  respect to the solar-scaled ratio may lead to similar observational
  results. We combine various TiO-based, IMF-sensitive indicators in
  the optical and NIR spectral windows, along with the FeH-based
  Wing-Ford band to break this degeneracy. We obtain a significant
  radial trend of the IMF slope in XSG1, a massive early-type galaxy
  (ETG), with velocity dispersion $\sigma\sim 300$\,\kms, observed
  with the VLT/X-SHOOTER instrument. In addition, we constrain -- for
  the first time -- both the shape and normalization of the IMF, using
  only a stellar population analysis.  We robustly rule out a single
  power-law to describe the IMF, whereas a power law tapered off to a
  constant value at low masses (defined as a bimodal IMF) is
  consistent with all the observational spectroscopic data {\sl and}
  with the stellar M/L constraints based on the Jeans Anisotropic
  Modelling method.  The IMF in XSG1 is bottom-heavy in the central
  regions (corresponding to a bimodal IMF slope \gammab$\sim$3, or a
  mass normalization mismatch parameter $\alpha\sim$2), changing
  towards a standard Milky-Way like IMF (\gammab$\sim$1.3;
  $\alpha\sim$1) at around one half of the effective radius. This
  result, combined with previous observations of {\sl local} IMF
  variations in massive ETGs, reflects the varying processes
  underlying the formation of the central core and the outer regions
  in this type of galaxies.
\end{abstract}


\section{Introduction}

Over the past few years, a number of papers have revisited the issue
of the non-universality of the stellar initial mass function (IMF) in
the unresolved stellar populations of early-type galaxies (ETGs). At
present, there are three independent probes of the IMF in the
unresolved, nearly quiescent populations of ETGs, based on the
modelling of the kinematics (e.g.~\citealt{Cappellari:2012a,
  Dutton:2013, Tortora:2013}); constraining the mass in strong
gravitational lenses that probe the projected mass distribution in the
central regions of galaxies
\citep[e.g.][]{Treu:2010,Auger:2010,Ferr:10}, or via gravity-sensitive
spectroscopic features (e.g.~\citealt{Cenarro:2003, vDC:10}). The
first two probes are sensitive to the stellar M/L, and their major
systematic relates to the contribution of the dark matter halo in the
central regions of galaxies. Spectroscopic constraints are mostly
sensitive to the ratio between giant and dwarf stars, and its main
systematic is the degeneracy with the additional parameters of the
underlying stellar populations, most notably the effect of variations
in the individual elemental abundances. We focus here on this issue.

A large effort has been made to understand this systematic in
detail. The bottleneck in the analysis is the lack of robust models of
stellar atmospheres, and their response to abundance effects. On this
front, population synthesis modellers follow either a theoretical
approach \citep[e.g.,][]{CvD12a} or an empirical methodology
\citep[e.g.,][]{Vazdekis:12}. The theoretical approach relies on
predictions from stellar atmosphere models to construct stellar
population models that best match the observed spectra, while the
empirical approach targets the deviation of the data from solar-scaled
stellar population models to measure the abundance effects.  At
present, both methods are needed to be able to improve our
understanding of line strengths in unresolved stellar populations. On
the issue of IMF variations, a number of papers have recently
strengthened the idea that ETGs feature a non-universal IMF
(see~\citealt{CvD12b,SPI:14}, as well as~\citealt[hereafter FER13 and
  LB13, respectively]{Ferr:13,LB:13}). Furthermore, the analysis has
gone as far as providing tentative drivers of IMF variations,
suggesting either [Mg/Fe] \citep{CvD12b}, velocity dispersion (FER13,
LB13,~\citealt{Cappellari:2013b}), metallicity \citep{MN:2015b}, or
more complex physical observables \citep{SPI:15b}. Studies combining
galaxy dynamics and spectroscopic features of the same galaxies have
revealed potentially serious systematics \citep{Smith:2014}, although
such effects could be strongly model-dependent
\citep{McDermid:2014,LB:15b}.  { In addition, constraints} based on
strong gravitational lensing are inconclusive: while
~\citet{Posacki:15} find agreement with the dynamical modelling,
several massive lensing ETGs feature stellar M/L values consistent
with a Kroupa-like IMF~\citep{Smith:2014}.  Critics have voiced
concerns on the interpretation of the line strength variations as an
IMF trend, proposing instead large departures from the solar-scaled
abundance of individual elements such as sodium
\citep{Jeong:13, Ziel:15}.  It is therefore imperative to understand
in detail the potential systematics (most notably, the effect of
elemental abundances, age, and metallicity). On the spectroscopic
analysis side -- the focus of this paper -- it is necessary to put
under the microscope several spectral features, ideally with varying
levels of dependence on the parameters under scrutiny.

{ The main target of the present work is a luminous ETG, 
hereafter XSG1, selected from the SPIDER survey~\citep{SpiderI},
more specifically from the subsample of $160$ ETGs with
$280<\sigma<320$~\kms\ defined by LB13.}
We have
gathered new, very deep, ``high'' resolution ($>4000$), long-slit
spectroscopy for XSG1 with the X-SHOOTER spectrograph on the ESO-Very
Large Telescope (VLT). The combination of depth, resolution, and
wavelength range for XSG1 gives us a unique opportunity to single out
the effect of the IMF against the other stellar population parameters
(in particular abundance ratios), as a function of galacto-centric
distance, advancing our previous
work~\citep[hereafter MN15a]{MN:2015a}.  To this effect, we focus on the optical,
TiO-based, spectral features, combining them with the Near-Infrared
TiO feature, \tionir , at $\lambda \sim$8860\,\AA, and the Wing-Ford
band, \feh\ ($\lambda \sim$9915\,\AA).  

The layout of the paper is as follows. In Sec.~\ref{sec:obs} we
describe the observations and instrumental setup. Sec.~\ref{sec:red} deals
with the data reduction. The analysis is detailed in
Sec.~\ref{sec:analysis}, including a description of the radial binning
of the data (Sec.~\ref{sec:kin}), the stellar population models used
(Sec.~\ref{sec:models}), the determination of stellar age
(Sec.~\ref{sec:age_z}), and abundance-ratio (Sec.~\ref{sec:profs_xfe})
profiles, as well as the method to constrain the IMF
(Sec.~\ref{sec:method}). Sec.~\ref{sec:res} presents the main results
of the work, i.e. the radial trends of IMF- and abundance-sensitive
features in XSG1, for TiO features (Sec.~\ref{sec:tio_profs}) and the
Wing-Ford band (Sec.~\ref{sec:feh}), the inferred radial trend of
IMF slope (Sec.~\ref{sec:gamma_trends}), as well as constraints to the
radial variation of IMF normalization, via a comparison of the stellar
population analysis with dynamics (Sec.~\ref{sec:alpha_trends}). A
discussion is presented in Sec.~\ref{sec:disc}, followed by a summary 
in Sec.~\ref{sec:summary}.

\begin{figure*}
\begin{center}
\leavevmode
\includegraphics[width=14cm]{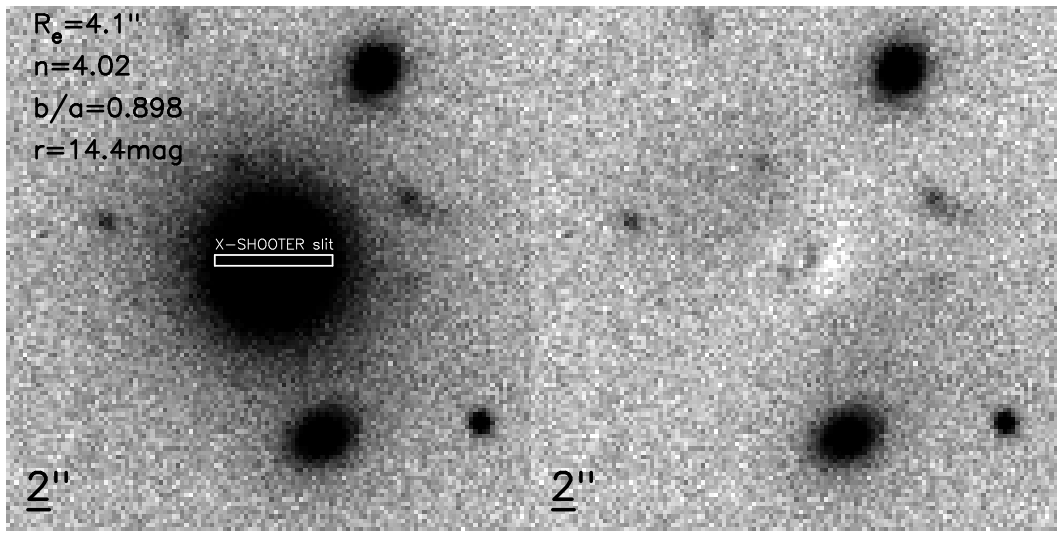}
\end{center}
\caption{SDSS $r$-band image of XSG1 ({\sl top--left}). The XSHOOTER
  slit is overplotted as a white rectangle.  North is right and East
  is up. The spatial scale (2'') is shown in the bottom-left corner.
  We include in this panel the effective radius, \re , the S\'ersic
  index, $\rm n$, the axis ratio $\rm b/a$, and the total $r$-band
  magnitude of the best-fitting S\'ersic model. {\sl Top-right:}
  residual map, after subtracting the best-fit S\'ersic
  model. 
  }
\label{fig:image}
\end{figure*}

\section{Observations}
\label{sec:obs}
The main target of the present work is XSG1 
(SDSS J142940.63+002159.0),
a massive ($\rm M_\star\,\sim\,2\times10^{11}M_\odot$; 
$\sigma\,\sim\,300$\,\kms ) ETG at redshift $z
\sim 0.057$, taken from the SPIDER survey~\citep{SpiderI}. The
galaxy has been selected from the sample of LB13 to have a high
abundance ratio (\afe$\sim 0.4$). Fig.~\ref{fig:image} shows the
SDSS r-band image of XSG1, together with the residual map obtained by
fitting the galaxy surface brightness distribution with a
two-dimensional, PSF-convolved, S\'ersic model, using the software {\sc
  2DPHOT}~\citep{LBdC08}. XSG1 is a round object ($\rm b/a \sim 0.9$),
whose light distribution is well described by a de Vaucouleurs
profile, with best-fitting S\'ersic $\rm n\,\sim\,4$, and an effective
radius of \re$\sim\,4.1$'' (see the Fig.~\ref{fig:image}), consistent with the SDSS
estimate of $\rm r_{e,SDSS} \sim 4.4$'' (based on a de Vaucouleurs
model).  

We have obtained new deep long-slit spectroscopy for XSG1
with the X-SHOOTER spectrograph at the ESO-VLT, on Cerro Paranal 
(Proposal IDs: 092.B-0378, 094.B-0747; PI: FLB).
X-SHOOTER is a second-generation ESO-VLT instrument -- a slit echelle
spectrograph that covers a wide spectral range (3000--25000\,\AA), at
relatively high resolution~\citep{Vernet:2011}. The incoming bin is
split into three independent arms (ultraviolet-blue, UVB:
3000--5900\,\AA; visible, VIS: 5300--10200\,\AA; near-infrared, NIR:
9800--25000\,\AA), each one equipped with optimized optics, dispersive
elements, and detectors, ensuring high sensitivity throughout the
entire spectral range.  Observations were done through five observing
blocks (OBs), each including two exposures on
target, interspersed by two sky exposures (through an
Object-Sky-Sky-Object sequence), with the same integration time as for the
science target.  For each OB, the on-target exposure time was 20.4,
22.6, and 25~min, resulting into a total exposure time of $\sim$1.7,
1.9, and $2.1$~hr, in the X-SHOOTER UVB, VIS, and NIR arms,
respectively. The observations were carried out in service mode during
two consecutive nights at the end of March 2014 (4 out of 5 OBs), and
completed at the end of February 2015. All data were taken under
photometric conditions, with seeing FWHM varying from $0.7$ to
$1.1$\,arcsec in the optical, with a mean value of $0.9$\,arcsec. The
X-SHOOTER slit is 11'' long, with a spatial scale of 0.16\,$\rm
arcsec/pixel$ in the UVB and VIS, and 0.21\,$\rm arcsec/pixel$ in the
NIR, arms. We adopted an instrument setup with 0.9''-, 0.9''-, and
1.0''- wide slits, providing a typical resolution power of 
$R \sim\!4400$, $\sim\!7500$, and $\sim\!5500$, in the UVB, VIS, and NIR arms,
respectively. The pointing was nodded by a few pixels along the slit
among exposures, in order to perform an optimal removal of cosmic rays
and detector defects (e.g. hot columns/pixels) during data
reduction. A hot (white dwarf) spectro-photometric standard star, together with a few
(3--7) telluric (B-type) standards were observed during each night
(the latter taken before and after each science OB) in order to perform
flux calibration and correct the spectra for telluric lines,
respectively.

As part of the same observing campaign as XSG1, we have also observed
a second target, a massive ($\rm M_\star\!\sim\!1.2\times 10^{11}M_\odot$; 
$\sigma\!\sim\!300$\,\kms ) ETG (hereafter XSG2; SDSS J002819.30-001446.7;
z=$\rm 0.059$; $\rm r_{e,SDSS}=4.1$'') also selected from the SPIDER
survey, with a more typical \afe\ ($\sim 0.25$~dex) for its velocity
dispersion.  XSG2 has been observed with XSHOOTER (Proposal ID:
094.B-0747; PI: FLB) through a series of 10 OBs, each consisting of an
Object-Sky-Object sequence. This different strategy, resulting into
half of the total integration time spent on sky, allowed us to perform
shorter individual exposures, in order to test the quality of sky
subtraction in the NIR, where the fast variability of sky lines may
hamper the analysis of the spectra, especially in the outer, low
surface-brightness regions. The data of XSG2 will be the subject of a
forthcoming paper. Here, we only use the central spectrum of XSG2 to
complement the analysis of the Wing-Ford band for XSG1, as detailed in
Sec.~\ref{sec:feh}.

\section{Data reduction}
\label{sec:red}

The echelle layout of X-SHOOTER produces 12--16 curved and highly
distorted spectral orders per arm, requiring a complex reduction
process. We refer the reader to~\citet{SCH:2014} for a detailed
description of all issues related to the data reduction of X-SHOOTER
UVB- and VIS-arm data. For our purposes, we performed the data
reduction using version 2.4.0 of the X-SHOOTER data-reduction
pipeline~~\citep{Mod:2010}, optimized with dedicated FORTRAN software
developed by the authors. We ran the X-SHOOTER pipeline in
``physical'' mode, through the ESO command-line utility {\sc
  esorex}. For each arm, the data (i.e. science and sky, as well as
standard star frames) were entirely pre-reduced with the X-SHOOTER
pipeline, including de-biasing, correction for dark frames (NIR-arm
only), rectification of echelle orders, wavelength-calibration, and
merging of the orders. The remaining reduction steps were performed as
follows:

\begin{description}
 \item[ {\it Flux calibration } - ] For the UVB and VIS arms, we found
   significant differences among reference spectra of standard stars
   to the flux-calibrated ones computed by the pipeline, being as high
   as 5--10$\%$ in regions affected by strong telluric lines (for the
   VIS arm).  For this reason, we applied an iterative procedure to
   improve the response function of the pipeline. The spectrum of each
   flux-calibrated standard star is corrected for tellurics with the
   same procedure as for the science data (see below).  Then, for each
   echelle order, we derive the ratio of the corrected standard-star
   spectrum to its tabulated version. The ratio is fitted,
   order-by-order, with Legendre polynomials of suitable orders (from
   one to twelve, depending on the spectral order), and the best-fitting
   polynomials are used to update the response function (for each
   observing night).  The procedure is iterated a few times, until a
   variation below 1\,$\%$ (the typical accuracy of our flux
   calibration) is achieved in the updated response function, at all
   wavelengths. For the NIR arm, the spectrum of the flux-calibrated
   standard-star, produced by the pipeline, turned out to have large
   deviations (up to $10-15\,\%$) with respect to the tabulated flux
   values. These deviations, with typical scales of $\sim 500$\,\AA, are
   likely due to residual sky-emission and telluric lines affecting
   the cubic-spline fit performed by the pipeline when deriving the
   response function. Therefore, we applied the flux-calibration of
   the NIR arm with a dedicated procedure. For each observing night,
   the (one-dimensional) non-calibrated standard-star spectrum is corrected for
   tellurics with the software {\sc molecfit } (see below). Intrinsic
   (mostly hydrogen) absorption lines, from the stellar atmospheres, are removed, by fitting them
   with a linear combination of multiple Gaussian functions. We
   compute the ratio of the resulting spectrum to the tabulated one by
   carefully masking out regions with strong telluric absorption
   ($>30\%$) and removing pixels with spurious spikes (e.g. from
   cosmic rays or residuals of sky lines), through a 2-$\sigma$
   clipping procedure.  The resulting ratio is fitted with a
   second-order spline to produce an initial response function. This
   response is applied to each telluric standard-star observed during
   the night. The continuum of the telluric standard is modelled, and
   removed, with two black-body (BB) laws, one for bluer and another
   one for redder wavelengths with respect to the Brackett jump
   ($\lambda\sim 16,000$\,\AA ). The temperatures of the two BBs are
   obtained by a best-fitting procedure. The resulting signal,
   median-smoothed, and averaged over all telluric-standards observed
   during the night, was applied as a correction to the initial
   response, to derive the final response function. Comparing the
   central spectrum of XSG1 among different exposures/nights, we have
   estimated that our NIR flux-calibration is accurate at $\simlt 2\,\%$.
 \item[ {\it Sky subtraction } - ] Our targets -- with an effective
   radius around $4$'' -- fill the whole X-SHOOTER 11''
   slit. Therefore, sky subtraction cannot be performed in the
   standard way of interpolating the sky from the two sides of the
   slit. We performed sky subtraction with the software {\sc
     skycorr}~\citep{Noll:2014}. Given the two-dimensional spectrum of
   the object, $O$, we consider the nearest sky frame, $S$. We extract
   a one-dimensional (1D) spectrum from both $S$ and $O$, by
   median-combining each frame over an 8-pixel region, using four
   pixels from each edge of the slit (where the signal from the galaxy
   is less prominent). We run {\sc skycorr} to determine the optimal
   scaling factors (depending on wavelength) that match the 1D
   spectrum of the sky to that of the object frames. The scaling
   factors are then applied to rescale $S$, and subtract it from $O$.
 \item[ {\it Telluric correction } - ] Telluric lines were removed from
   each object exposure individually. For the VIS arm, we adopted two
   independent approaches, by (i) using telluric standard stars
   observed during our programme, and (ii) using the software {\sc
     molecfit}~\citep{Smette:2015, Kausch:2015}. In case (i), we
   obtained, for each telluric standard star, an atmospheric
   transmission curve, by removing intrinsic absorption lines and the
   continuum from the stellar spectrum. The correction itself was
   performed by deriving the linear combination of (re-scaled)
   transmission curves that minimizes the absolute deviation to the
   object spectrum, computed over spectral regions with prominent
   telluric lines. In case (ii), {\sc molecfit} was run to compute a
   theoretical transmission model from the object spectrum itself. In
   general, both approaches were found to provide very consistent
   results (with differences below $1\%$), as also shown in the analysis of
   line strengths throughout the present work.  For the NIR arm, given
   the excellent quality of {\sc molecfit} fits, we decided to rely
   only on this method.
 \item[ {\it Combination } - ] To combine different object exposures,
   we first aligned the photometric centre of the galaxy as a function
   of wavelength in each two-dimensional spectrum, in order to remove 
   its spatial drift due to the atmospheric dispersion. To this effect, we
   collapsed the object spectrum along the slit direction in wavelength
   bins (with size $\sim 100$\,\AA ), and derived the photometric
   centre, for each bin, by fitting the collapsed profile with a
   linear combination of Gaussian functions. The aligned 2D spectra
   were then averaged with the {\sc IRAF} tool {\it imcombine}. Cosmic
   rays, as well as pixels flagged by the X-SHOOTER pipeline were
   excluded from the average. A noise map was propagated accordingly.
\end{description}


\section{Analysis}
\label{sec:analysis}

\begin{figure}
\begin{center}
\leavevmode
\includegraphics[width=8.2cm]{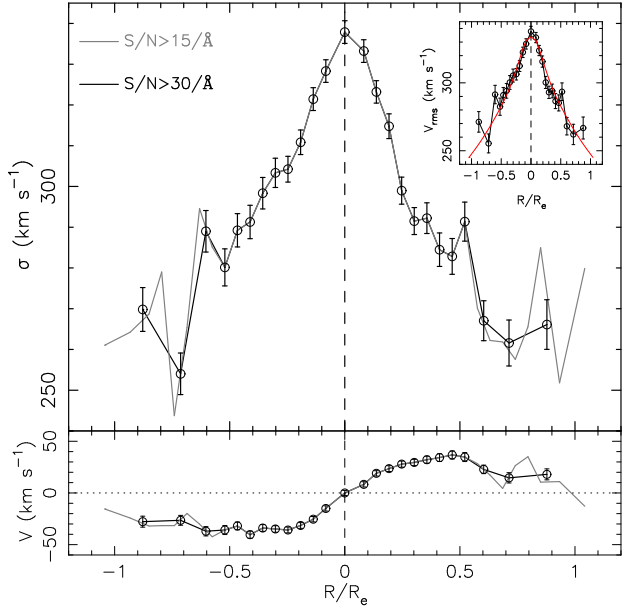}
\end{center}
\caption{Velocity dispersion (top) and rotation velocity (bottom)
  profiles, obtained with {\sc pPXF}, for XSG1, as a function of
  galactocentric distance, $\rm R/R_e$. Black and grey curves
  correspond to radial binned spectra, along the X-SHOOTER slit, with
  minimum S/N ratio of 30 and 15, respectively. Error bars
  denote $1\sigma$ uncertainties. Notice that the galaxy has low
  rotation, with a radial drop in velocity dispersion (from $\sim 340$\,\kms\ in
  the centre, to $\sim 270$\,\kms\ at 1\,\re ). The inset in the top panel
  compares the observed $\rm V_{rms}=\sqrt{V^2+\sigma^2}$ (empty
  circles) with best-fitting results from the Jeans Anisotropic
  modelling method \citep[JAM, red curve in the
   inset;][]{Cappellari:2008}.  }
\label{fig:kin}
\end{figure}

\begin{figure*}
  \raisebox{-0.5\height}{\includegraphics[width=8cm]{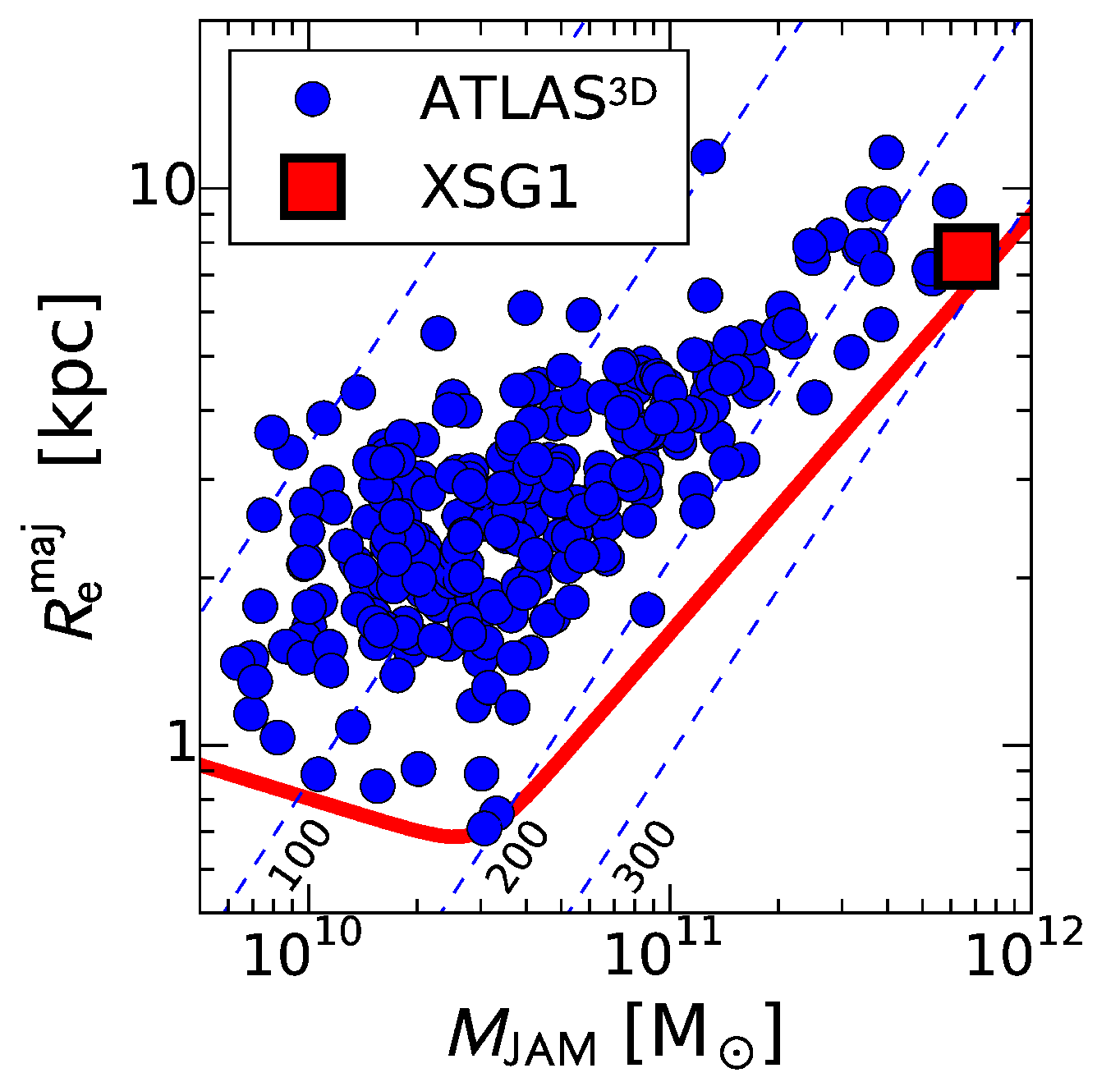}}
  \raisebox{-0.5\height}{\includegraphics[width=8.5cm]{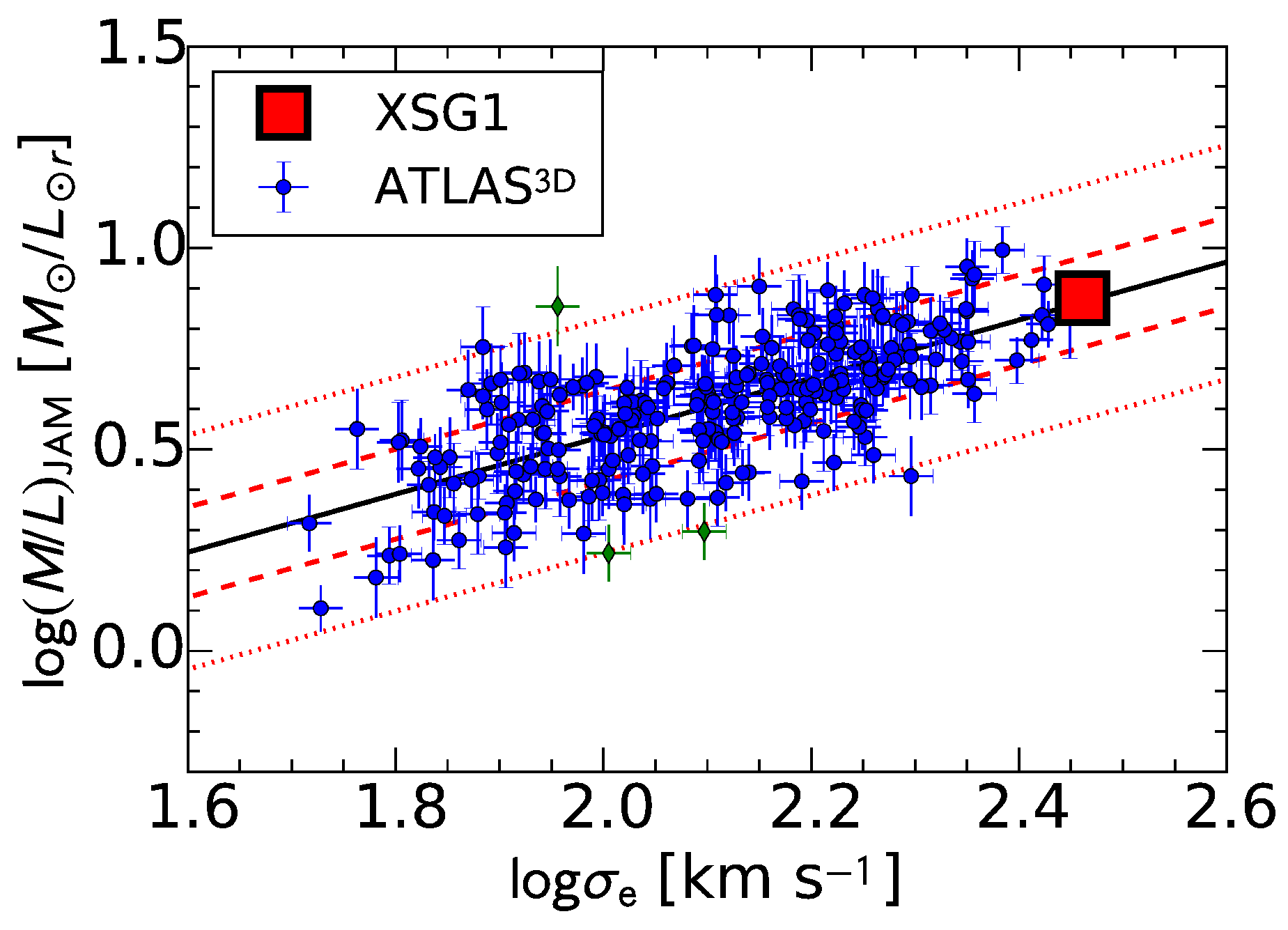}}
  \caption{The early-type galaxy XSG1 (filled red square) is compared
    with the ATLAS$^{\rm 3D}$ sample \citep[blue
      dots,][]{Cappellari:2013} by use of the Jeans Anisotropic
    Modelling (JAM). {\sl Left:} On the size-mass plane XSG1
    represents a typical ETG at the massive end of the distribution.
    The dashed lines mark contours of constant velocity dispersion, as
    labelled (in km/s). The thick red line is the zone of exclusion,
    as given by eq.~4 in \citet{Cappellari:2013b}. Note that the
    effective radius here is measured, for consistency, following the
    same methodology as in the ATLAS$^{\rm 3D}$ sample (see text for
    details). {\sl Right:} The { stellar r-band M/L} derived from the JAM
    modelling makes XSG1 a typical massive ETG. The ($1\sigma$) error
    bars for XSG1 are smaller than the size of the square. The dashed
    and dotted red lines indicate the $1\sigma$ and $2.6\sigma$ (99\%)
    observed scatter around the best-fitting relation of the
    ATLAS$^{\rm 3D}$ sample (black solid line).  }
  \label{fig:jam}
\end{figure*}

\subsection{Kinematics, JAM modelling, and radial binning}
\label{sec:kin}

\subsubsection{Kinematics}
In order to compute the kinematics of XSG1, we binned the X-SHOOTER
two-dimensional spectrum along the spatial direction, starting from
the photometric centre of the galaxy.  We chose radial bins with a
minimum width of 0.22'' (i.e.  about 1/4 of the seeing FWHM), and
increased the bin width adaptively to reach a given, minimum (median)
signal-to-noise ratio, $\rm S/N_{min}$ (computed per \AA, at $\lambda
\lambda\!\sim\! 4800-5600$\,\AA ), in each bin. This procedure allowed
us to reach an outermost radius of $\sim 0.88$ ($1.05$)~\re\ for $\rm
S/N_{min}=30$ ($15$). For each radial bin, we measured the recession
velocity and the galaxy velocity dispersion using the software
\textsc{pPXF}~\citep{Cap:2004}, { which simultaneously fits the
  stellar kinematics and an optimal linear combination of spectral
  templates to the observed spectrum, using a maximum-likelihood
  approach that works in wavelength space,.}  We measured the kinematics
independently in different spectral regions of the UVB and VIS arms
($\lambda\lambda =4000-9000$\,\AA ), combining the corresponding
probability distribution functions into final estimates. { From the
  recession velocity } in the innermost bin,
we obtain a redshift of z=$0.05574\pm0.00001$ for XSG1, fully consistent with the value of
z=$0.05575\pm0.00001$ from the SDSS database. Fig.~\ref{fig:kin} plots
the profiles of rotation velocity, $\rm V$, and velocity dispersion,
$\sigma$, for XSG1, showing that the galaxy is likely a
pressure-supported system, with $\sigma$ decreasing smoothly from
$\sim 340$\,\kms\ in the centre, to $\sim 270$\,\kms\ at 1\,\re , along
both sides of XSG1. The galaxy rotation velocity is less than $\sim
50$~\kms\ at all radii. Notice that we get very consistent kinematics
when binning the spectra with either $\rm S/N_{min}=30$ or $15$, as shown
by the black and grey lines.  Under the assumption of
circular symmetry, the velocity dispersion profile in
Fig.~\ref{fig:kin} implies a luminosity-weighted value $\sigma\sim
317$\,\kms\ within the SDSS fibre aperture radius of 1.5'', consistent
with two estimates of $\sigma = 301 \pm 9$ and $313 \pm 10$\,\kms\ 
available from the SDSS database.

\begin{table}
\centering
\small
 \caption{Relevant properties of radially binned spectra used to
   constrain the IMF.  For all bins we fold-up the spectra around the
   photometric centre of the galaxy.  }
  \begin{tabular}{c|c|c|c}
   \hline
 Radius & Radial range & $\sigma$ & $\rm S/N$ \\
  ('')  &    ('')      &  (\kms ) &           \\
   (1)  &     (2)      &   (3)    &    (4)     \\
  \hline
$0.    $  &  $[-0.675,+0.675]$ &  $333\pm3$  &  $320$ \\
$0.7875$  &  $[0.675,0.900]$   &  $316\pm3$  &  $130$ \\
$1.025 $  &  $[0.900,1.150]$   &  $307\pm3$  &  $110$ \\
$1.3250$  &  $[1.150,1,500]$   &  $302\pm3$  &  $100$ \\
$1.7750$  &  $[1.500,2.050]$   &  $291\pm3$  &   $90$ \\
$3.1250$  &  $[2.050,4.200]$   &  $276\pm3$  &   $90$ \\
   \hline
  \end{tabular}
\label{tab:bins}
\end{table}

\begin{figure*}
\begin{center}
\leavevmode
\includegraphics[width=18cm]{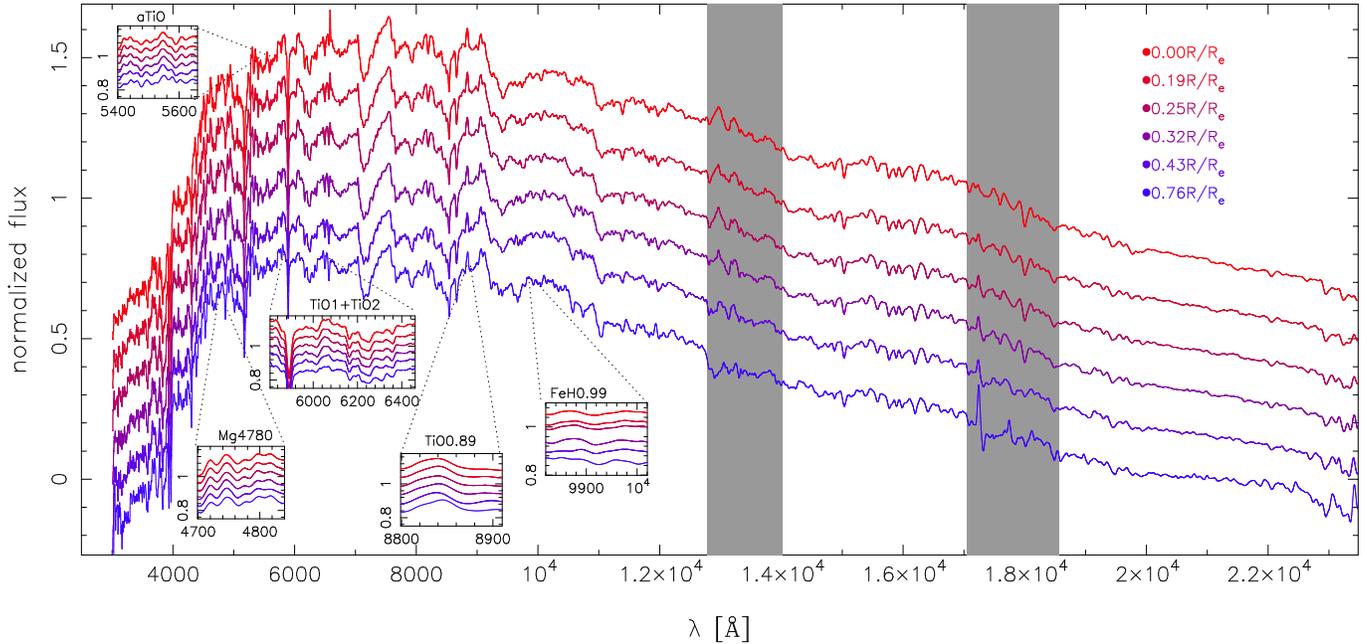}
\end{center}
\caption{X-SHOOTER spectra of XSG1 used to constrain the IMF. The data
  are radially binned in six galacto-centric distances, out to $\rm
  R/R_e \sim 0.8$. Different colours, from red through blue,
  correspond to different radial positions, as labelled. The insets
  are zoomed-in versions of the main spectral features used in the present work
  (see Sec.~\ref{sec:method}).  All spectra have been smoothed, for
  displaying purposes, to a common velocity dispersion of 400\,\kms .  }
\label{fig:spectra}
\end{figure*}

\subsubsection{JAM modelling}
The inset in the top panel of Fig.~\ref{fig:kin} compares the observed
RMS velocity, $\rm V_{rms}=\sqrt{V^2+\sigma^2}$ (see empty circles),
with best-fitting predictions from the spherical version of the Jeans
anisotropic modelling (JAM, red curve in the
inset;~\citealt{Cappellari:2008}).  The use of a spherical model is
motivated by the fact that above the critical mass $M \sim 2 \times
10^{11} \, M_\odot$, ETGs are fully dominated by slow rotator
galaxies, which are intrinsically quite close to
spherical~\citep{Cappellari:2013b}.  However, we also verified that,
unlike the spherical model, an axisymmetric, intrinsically flat, but
nearly face-on JAM model, could not well describe the observed $\rm
V_{rms}$ profile with a constant anisotropy, as one generally finds
for fast rotators.  Finally, from the $\rm {V_{max}}/\sigma_0$ value
computed from the long-slit spectrum, converted into $\rm
(V/\sigma)_e$ using eq.~(23) of~\citet{Cappellari:2007}, the galaxy is
securely classified as a slow rotator.

The JAM model is based on a Multi-Gaussian Expansion (MGE;
~\citealt{Emsellem:1994}) model of the { SDSS r-band image}, using
the Python method and software of~\citet{Cappellari:2002}.  The JAM
model assumes mass follows light~\footnote{ Notice that although the
  assumption of (total) mass following light might be inconsistent
  with the finding that the stellar M/L varies with radius (see
  Sec.~\ref{sec:alpha_trends}), as shown in figure~9
  of~\citet{Cappellari:2013}, the $\rm (M/L)_{JAM}$ is actually robust
  against different assumptions, such as the inclusion of a more
  realistic (NFW) dark-matter halo, in the modelling.  }, within the
region sampled by the kinematics, which implies any possible dark
matter within about 1~$\rm R_e$ is accounted for by a change in the
total $\rm (M/L)_{JAM}$. However, the dark-matter fraction is expected
to be $\la 30 \%$ in the mass range ($2 \times 10^{11} \, M_\odot$) of
XSG1~\citep{Cappellari:2013}.  The model adopts a supermassive
black-hole (SMBH) mass from the \citet{KormendyHo:2013} SMBH-$\sigma$
relation, and has a best-fitting radial anisotropy $\rm
\beta_r=0.16$. This small and positive radial anisotropy is typical of
slow rotators in the stellar mass range of XSG1
\citep{Cappellari:2007}. The best-fitting total $\rm (M/L)_{JAM}$,
{ computed in the SDSS-$r$ band}, is $7.50 \pm 0.75$. The
error bar on $\rm (M/L)_{JAM}$ is not just a formal error but is a
quite conservative estimate, accounting for possible systematics in
the modelling.  Notice that the SMBH value is not critical in the
derivation of $(M/L)_{JAM}$, but not completely negligible either. A
model without a SMBH would give nearly the same $\rm (M/L)_{JAM}$, but
would require a larger anisotropy.

Fig.~\ref{fig:jam} compares XSG1 with respect to the local sample of
ETGs from the ATLAS$^{\rm 3D}$ survey \citep{Cappellari:2013b}. The
position of our galaxy both on the mass-size plane ({\sl left panel})
and the total (M/L) versus $\rm \sigma_e$ relation ({\sl right}) is
representative of a typical massive ETG.  Notice that for this
analysis, we follow, for consistency, the same procedure adopted by
~\citet{Cappellari:2013}. In particular, we calculate the $\rm
R_e=5.17''$ as the radius enclosing half of the analytic MGE total
light. This value is then multiplied by the factor $1.35$ to bring it
into consistency with the RC3 system (see figure~7
of~\citealt{Cappellari:2013}).  This scaling factor is not universal,
but it is justified by the facts that we are using the same kind of
photometry as in the ATLAS{\rm 3D} work.  For a flat Universe
($\Lambda$CDM with $\Omega_m=0.3$ and
$H_0=70$\,km\,s$^{-1}$\,Mpc$^{-1}$) this result maps into a physical
effective radius of $\rm R_e=7.6$\,kpc. Note this is the value
adopted, for consistency, when comparing with the ATLAS$^{\rm 3D}$
sample (Fig.~\ref{fig:jam}), in contrast with $\rm R_e=4.1$''
(4.4\,kpc) derived in Sec.~\ref{sec:obs} from a S\'ersic fit to the
SDSS-$r$ band surface brightness profile. Note that in the rest of the
paper, we use the latter as the fiducial value for $\rm R_e$.

\subsubsection{Radial binning for stellar population analysis}

To constrain the IMF at different galacto-centric distances, we
construct six radially binned spectra, folding up the spectra from the
opposite sides of the slit around the photometric centre. To this
effect, we offset each row in the two-dimensional spectrum to the
rest-frame, adding up the spectra in radial bins. The bin width is
increased adaptively, if needed, in steps of $0.225$'' (1/4 of the
seeing FWHM), targeting a $\rm S/N>90$ per radial bin. In order to
minimize seeing effects, the innermost spectrum has a width of 1.3''
(i.e. $\pm 0.675$'') around the photometric centre, corresponding to a
factor of 1.5 times the mean seeing FWHM of our data. The outermost
bin reaches an average galacto-centric distance of $\sim$2/3\,\re
. Since the velocity dispersion is one of the main parameters to
interpret stellar populations via line-strengths, we have re-run 
{\sc pPXF} on the six binned spectra to get an accurate velocity
dispersion estimate for each bin. The main characteristics of the
spectra, i.e.  radial range, average galacto-centric distance, median
$\rm S/N$, and $\sigma$ are summarized in
Tab.~\ref{tab:bins}. Fig.~\ref{fig:spectra} shows the six binned
spectra of XSG1, zooming into those regions used to constrain the IMF in the
present work.

\begin{table*}
\centering
\small
 \caption{Targeted spectral indices applied to constrain the
   IMF. Wavelengths are quoted in the air system.  Notice that the
   \tionir\ index has no units, as it is defined as a flux ratio between
   the blue and red passbands.}
  \begin{tabular}{c|c|c|c|c|c}
   \hline
 Index  & Units & Blue Pseudo-continuum & Central feature & Red Pseudo-continuum & Ref. \\
        &       & [\AA] & [\AA] & [\AA] &      \\
   (1)  &   (2) &   (3) &   (4) &   (5) & (6)  \\
   \hline
\mgf    & \AA & $4738.9$--$4757.3$ & $4760.8$--$4798.8$ & $4819.8$--$4835.5$ &             Serven+05   \\
\tioi   & mag & $5816.625$--$5849.125$ & $5936.625$--$5994.125$ & $6038.625$--$6103.625$&  Trager+98   \\
\tioii  & mag & $6066.625$--$6141.625$& $6189.625$--$6272.125$ & $6422.0$--$6455.0$&       $\rm LB13$  \\
\atio   & mag & $5816.625$--$5849.125$ & $5936.625$--$5994.125$ & $6038.625$--$6103.625$ & $\rm STK14$ \\
\tionir &     & $8832.569$--$8852.563$ &  & $8867.559$--$8887.554$ & $\rm CvD12$ \\
\feh    & \AA & $9852.292$--$9877.285$ & $9902.278$--$9932.270$ & $9937.268$--$9967.260$ & $\rm CvD12$  \\
  \hline
  \end{tabular}
\label{tab:indices}
\end{table*}

\subsection{Stellar population models}
\label{sec:models}


We analyze the X-SHOOTER spectra with extended MILES stellar
population models~(\citealt{RV:15}; R\"ock et al., 2015b, A\&AR, in
preparation), the first set of single-burst stellar population models
to date that cover both the optical and infrared wavelength range,
between 3500 and 50000\,\AA, based exclusively on empirical stellar
spectra. These models combine MIUSCAT stellar population models
(\citealt{Vazdekis:12, Ricciardelli:2012}), covering the spectral
range $\lambda\lambda 3465-9469$\,\AA\ at a nominal resolution of
$2.51$\,\AA\ FWHM~\citep{Jesus:11}, with the NIR model extension based
on the IRTF stellar library, covering the $\lambda\lambda
8150-50000$\,\AA\ wavelength range at a spectral resolution of
2000~\citep{RV:15}. The MIUSCAT and IRTF models are joined in the
interval $\lambda\lambda 8950-9100$\,\AA . For $\lambda<8950$~\AA ,
the model predictions are identical to MIUSCAT. The extended MILES
models rely on solar-scaled isochrones with stellar spectra following
the abundance pattern of our Galaxy, i.e.  approximately solar-scaled
at solar metallicity, and are derived from either Padova00 or BaSTi
isochrones. For comparison with our previous works (e.g.~FER13; LB13;
MN15a), we base our analysis, throughout the present work, on Padova
models. Because of the IRTF implementation, the extended MILES SSPs
cover a restricted parameter range than MILES models, i.e. ages from
$1.0$ to $17.78$\,Gyr, and four bins of total metallicity, i.e.  $\rm
[M/H]=\{$ $-0.71$, $-0.4$, $0$, $+0.22\}$.  The SSPs are provided for
several IMFs, as for MILES models, i.e. unimodal (single power-law)
and bimodal (low-mass tapered) IMFs, both characterized by their
slope, $\Gamma$ (unimodal) and $\rm \Gamma_b$ (bimodal), as a single
free parameter (see, e.g.,~\citealt{Vazdekis:1996, Vazdekis:2003,
  Ferreras:2015}). { The bimodal IMFs are smoothly tapered off
  below a characteristic ``turnover'' mass of $0.6$\,M$_\odot$.}
For $\Gamma_b \sim 1.3$, the bimodal IMF gives a good representation
of the Kroupa IMF, while for $\Gamma\sim 1.3$ the unimodal IMF
coincides with the \citet{salp:55} distribution.  The lower and upper
mass-cutoff of the IMFs are set to $0.1$ and $100$\,M$_\odot$,
respectively.  Notice that younger ages, as well as lower
metallicities, than those provided by extended MILES models are not
relevant to the study of massive ETGs, while a mild extrapolation of
the models to higher metallicity (from $\rm [M/H]\sim +0.22$ up to
$\sim +0.25$) is required to match the innermost spectrum of XSG1 (see
Sec.~\ref{sec:age_z}).  As in FER13 and LB13, we consider bimodal IMF
models with $\Gamma_b=\{0.3$, $0.8$, $1.0$, $1.3$, $1.5$, $1.8$,
$2.0$, $2.3$, $2.8$, $3.3\}$, while for unimodal IMFs, we restrict the
analysis to $\Gamma \le 2.3$, where model predictions are safe
(see~\citealt{Vazdekis:12}).

To estimate the effect of abundance ratios on line strengths, we also
combine the predictions of extended MILES models, with those from other two sets
of models, i.e. $\alpha$--MILES~\citep{Vazdekis:15}
and~\citet[hereafter CvD12]{CvD12a} SSPs. The \amiles\ SSPs, 
{ having a spectral resolution of 2.51~\AA\ (FWHM), are based
on the MILES library and apply corrections from theoretical models of
stellar atmospheres to produce  spectra of old- and intermediate-age
stellar populations } with varying total
metallicity, IMF, and \afe\ abundance ratios. The models are based on
BaSTI scaled-solar and $\alpha$--enhanced isochrones. The CvD12 models
combine optical (MILES) and near-IR (IRTF) empirical stellar libraries
to produce integrated light spectra, at $\lambda\lambda=0.35-2.4$\,$\mu$m, 
with a resolving power of $R \sim 2000$.  The empirical CvD12
models, computed at solar metallicity and in the age range from $3$ to
$13.5$\,Gyr, are complemented with synthetic SSPs modelling the
spectral variations due to changes in individual elemental abundances,
as well as in generic $\alpha$ elements. We notice that CvD12 models
are computed at fixed $\rm [Fe/H]$, rather than $\rm [M/H]$ (adopted in the
extended- and $\alpha$-MILES models), and based on a combination of three
different sets of scaled-solar isochrones. We also emphasize that in
the publicly available version of CvD12 models, SSPs with varying
elemental abundances are only provided for a specific age ($13.5$\,Gyr)
and IMF (Chabrier), and $\rm [Fe/H]=0$. Because of the difference
between \amiles\ as well as CvD12 models and our reference extended MILES
SSPs (i.e. different isochrones, metallicity range, etc.....), we use
these other models mostly to perform a qualitative analysis of the
data, whereas the quantitative results presented in this work
(e.g.~Sec.~\ref{sec:gamma_trends}) are entirely based on extended MILES
models.

Notice that the present work relies on the analysis of line-strengths,  based on 
several index-index diagrams (see Sec.~\ref{sec:res}). In these diagrams -- mostly used for illustrative purposes --  
all model line-strengths are computed after smoothing the models to a $\sigma$ of 300~\kms\ (accounting for the intrinsic 
spectral resolution of the models), i.e. the averaged velocity dispersion (among radial bins) for XSG1. 
In all index-index diagrams, observed line-strengths are corrected to a common $\sigma$ of 300~\kms . 
For a given radial bin, the correction is obtained using the corresponding best-fit extended-MILES SSP (see Sec.~\ref{sec:gamma_trends}). We compute the difference of each line-strength when smoothing the model to the actual 
$\sigma$ of the spectrum and the reference value of 300~\kms .
Given the relative narrow range in velocity dispersion for XSG1, and the fact that we mostly analyze broad features 
(e.g. TiO and \feh ), the correction is largely independent of the specific model (e.g. age and metallicity)
adopted. 
However, in our quantitative analysis (i.e. when fitting model to observed line-strengths), we compute, for each bin,
line-strengths at the original $\sigma$ of the data, and smooth the models accordingly. We prefer this approach, rather than
smoothing all observed spectra to the same $\sigma$, as it extracts the maximum amount of information 
from the data (see, e.g., LB13), and avoids any contamination of the relevant features, when smoothing the spectra, from nearby sky residuals.

\begin{figure}
\begin{center}
\leavevmode
\includegraphics[width=8.2cm]{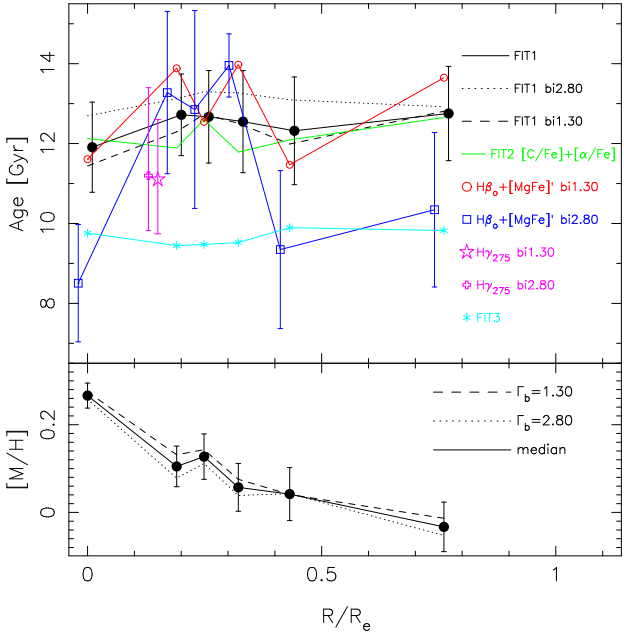}
\end{center}
\caption{Radial trends of age ({\sl top}) and metallicity ({\sl
    bottom}) in XSG1. Different symbols and line types correspond to
  different methods to derive age and metallicity, as labelled (see
  text for details). Error bars denote $1\sigma$ uncertainties.  }
\label{fig:ageZ}
\end{figure}

\begin{figure}
\begin{center}
\leavevmode
\includegraphics[width=8.2cm]{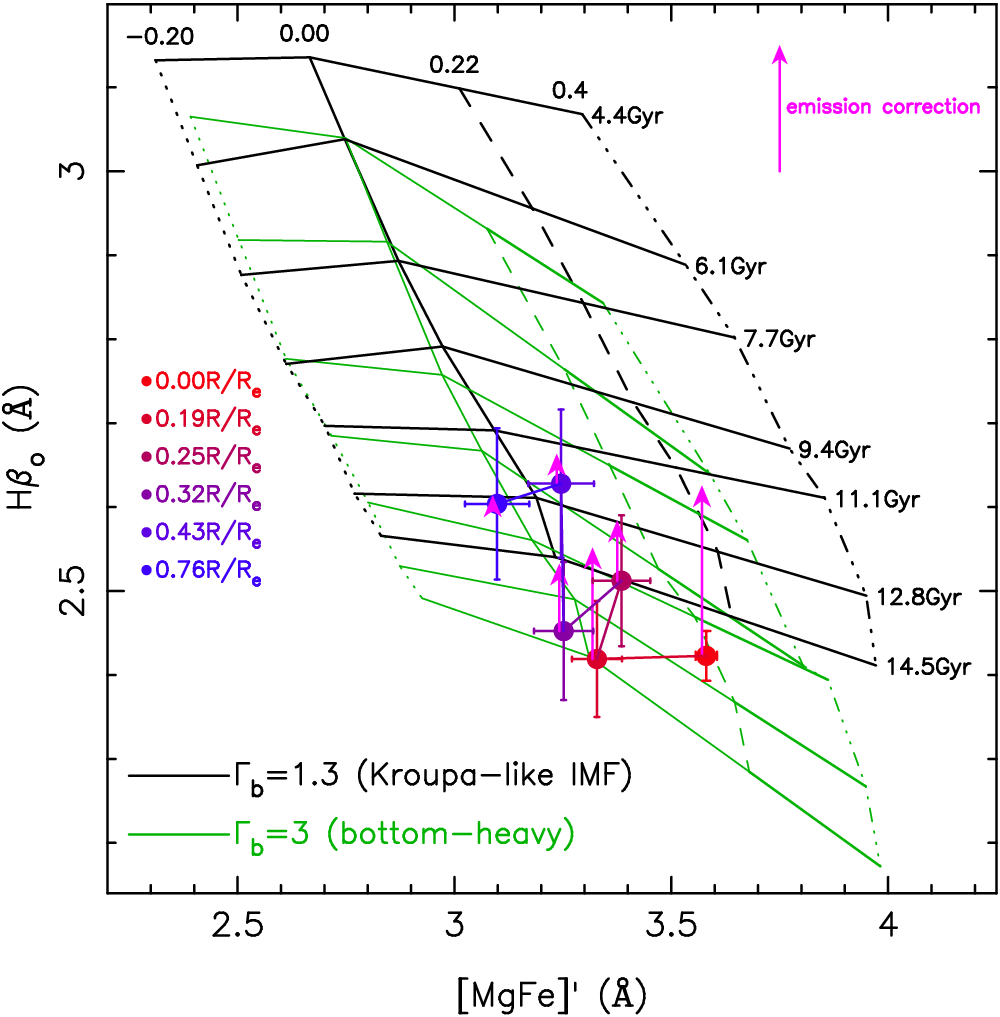}
\end{center}
\caption{The age indicator \hbo\ is plotted as a function of the total
  metallicity index \mgfep . The solid dots represent the observed
  line-strengths in XSG1 ($1\sigma$ error bars), corrected to a common
  velocity dispersion of 300\,\kms\ (see Sec.~\ref{sec:models}). The
  black and green grids correspond to Kroupa-like (\gammab$=1.3$) and
  bottom-heavy (bimodal IMF; \gammab=$3$) MILES model predictions
  with varying age and metallicity, as labelled.  Notice that, for the
  specific purpose of this plot, the grids have been linearly
  extrapolated from the maximum value of \zh$=+0.22$ in MILES
  models, up to \zh$=+0.4$. Different radial bins are shown
  (inside-out) with red-through-blue colours (see labels on the left
  side of the plot). The effect of emission line correction on \hbo\ is
  shown with magenta arrows.  }
\label{fig:hbo}
\end{figure}

\subsection{Radial profiles of age and metallicity}
\label{sec:age_z}

An accurate estimate of age and metallicity is needed to constrain the
IMF, because gravity-sensitive spectral features also depend, to some
extent, on such parameters. We estimate radial profiles of age and
\zh\ for XSG1 with different methods, as illustrated in
Fig.~\ref{fig:ageZ}.  In particular, regarding stellar age, we explore
several approaches (see labels in the top panel of the Figure):
\begin{description}
\item[ {\it FIT1} - ] The age is derived by performing spectral
  fitting over the wavelength range
  $\lambda\lambda\!=\!3800-6200$\,\AA .  The fit is done individually
  for extended MILES SSPs with different choices of \gammab \, (see Sec.~\ref{sec:models}),
  normalizing both model and observed spectra by the median flux in a
  { reference} spectral window ($\lambda\lambda=5200-5400$\,\AA ).  The black,
  solid line, in the Figure plots the median age value, for each
  radial bin, among SSPs with different \gammab . The black error bars
  reflect the statistical uncertainty on the age, summed in quadrature
  to the standard deviation among estimates for different values of
  the IMF slope, \gammab. We find no significant age gradient
  throughout the galaxy. This result is further confirmed when fitting
  models with extended star-formation histories to the spectra (see
  FER13 and LB13 for details), in which case we find very small
  differences ($\simlt 0.2$\,Gyr) between luminosity- (as well as
  mass-) weighted ages with respect to the results with SSP models,
  meaning that at all radial bins the spectra of XSG1 are very well
  described by a single SSP. This is consistent with the high
  \afe\ (implying a very short star-formation time-scale in the galaxy
  within the effective radius, see below). The dotted and dashed black
  lines in the top panel of Fig.~\ref{fig:ageZ} show the results for
  a Kroupa-like (\gammab$=1.3$) and a bottom-heavier (\gammab$=3$)
  IMF, showing that, for our X-SHOOTER data, the age profile does not
  depend on \gammab . On the other hand, we found that
  the absolute value of the age depends on the
  adopted methodology, e.g. the way one normalizes the continua of
  observed and model spectra prior to spectral fitting. The cyan
  profile in the Figure (labelled $FIT3$) shows the case when one
  divides the spectra by their integrated flux in the normalization
  window (see above), rather than the median flux (as for the solid black curve).
  Although the age gradient does not change, we find an overall shift
  of $\sim$2\,Gyr in the age values.  In general, changing the setup
  of the spectral fitting approach (e.g. the continuum normalization
  window), we found variations in the mean age from $10$ to $13$\,Gyr.
\item[ {\it FIT2 \, $[C/Fe]+[\alpha/Fe]$} - ] The solid green line in
  the top panel of Fig.~\ref{fig:ageZ} shows the spectral fitting
  constraints when ``removing'' the effect of \cfe\ and
  \afe\ abundance ratios from the observed galaxy spectrum at each
  radial bin. In practice, for each radial bin, we use CvD12 SSP
  models to estimate the ratio of \cfe- and \afe- enhanced SSPs with
  respect to the solar-scaled SSP, and divide the observed spectrum by
  this ratio.  The ratio is estimated using the median values of
  \cfe\ and \afe\ measured for each bin
  (Sec.~\ref{sec:profs_xfe}). Notice that we consider here only the
  effect of \cfe\ and \afe\ as (i) both turn out to be enhanced in
  XSG1, (ii) they are the leading abundances to affect the TiO-based
  optical features analyzed in the present work (see
  Sec.~\ref{sec:tio_profs}), and (iii) CvD12 models suggest that
  \afe\ and \cfe\ are the main abundance ratios whose enhancement 
  affects the shape of the continuum, possibly leading to a biased
  determination of age from spectral fitting. Notice that enhancing
  \afe\ tends to make a given spectrum bluer, mimicking the effect of a
  young age (see, e.g.,~\citealt{Vazdekis:15}). On the contrary,
  enhancing \cfe\ tends to counteract this effect (i.e. reddening the
  spectrum; see~\citealt{Lee:2010}). As a result, for XSG1, accounting
  for the effect of abundance ratios does not change the
  age determination with respect to solar-scaled models.
\item[ {\hbo } - ] We use predictions from MILES SSP models to fit
  simultaneously the \hbo\ age-sensitive
  indicator~\citep{Cervantes:2009} and the total metallicity indicator
  \mgfep~\citep{Thomas:03}.  The EWs of \hbo\ are corrected for
  emission contamination as described in LB13, i.e. estimating the
  excess of flux in the line with respect to a combination of two SSPs
  that gives the best fit in the \hb\ spectral region ($\lambda\lambda
  = 4530-4730$\,\AA ) after excluding the trough of the absorption.
  The correction to \hbo\ varies from $\sim 0.2\,$\AA\ in the
  innermost bin, to less than $\sim 0.05$\,\AA\ in the outermost
  regions. Fig.~\ref{fig:hbo} shows the radial variation of \hbo\ and
  \mgfep\ line-strengths in XSG1. The corrected values of \hbo\ suggest
  no significant radial gradient in XSG1, as confirmed by the age
  trend inferred for a Kroupa-like IMF from the \hbo --
  \mgfep\ diagram (red solid line in top panel of
  Fig.~\ref{fig:ageZ}). Notice, however, that MILES models predict a
  decrease of \hbo\ with IMF slope, as seen by comparing green
  vs. black grids in Fig.~\ref{fig:hbo}, corresponding to models with
  \gammab=3 and \gammab=1.3, respectively (see LB13 and
  \citealt{Vazdekis:15} for more details). As a result, for a
  bottom-heavy IMF, one infers younger ages than for a Kroupa-like
  distribution (see blue solid line in top panel of
  Fig.~\ref{fig:ageZ}).
\item[ {\hgs} - ] By summing up the three innermost galaxy spectra of
  XSG1, the median $\rm S/N$ in the $\rm H\gamma$ spectral region
  ($\lambda \! \sim \! 4340$\,\AA ) is high enough ($>350$) to
  constrain the age with the {\hgs} spectral index, proposed
  by~\citet{VazA:1999} (see also~\citealt{Yamada:2006}). This 
  index is a pure age indicator, i.e.  it
  is completely independent of metallicity (provided that the wavelength
  calibration is accurate enough, as it is the case for our X-SHOOTER
  data). The magenta symbols in the top panel of Fig.~\ref{fig:ageZ}
  show the age inferred from \hgs\ when using either Kroupa-like
  (\gammab=1.3) or bottom-heavy (\gammab=3) SSP models.  The
  age inferred in the galaxy centre from \hgs\ ($\sim 11$\,Gyr) is
  independent of the IMF, and is consistent with those obtained from
  spectral fitting ($\sim 10-12$\,Gyr, see black and cyan lines in
  the Figure).
\end{description}

\begin{figure}
\begin{center}
\leavevmode
\includegraphics[width=8.2cm]{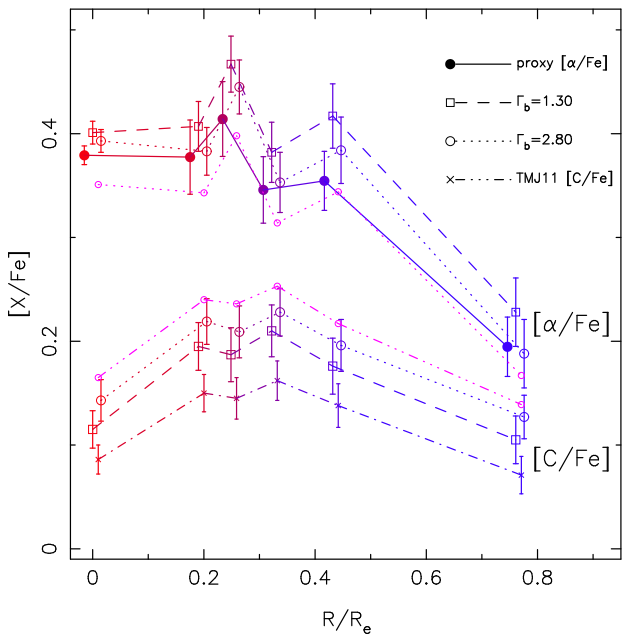}
\end{center}
\caption{Radial trends of elemental abundance ratios in XSG1.  We plot
  \cfe\ and \afe\ abundance ratios. These are expected to contribute
  the most to variations in the TiO features (see text, and
  Fig.~\ref{fig:tio_nocorr}).
The upper (lower), red-through-blue, curves plot the \afe\ (\cfe )
radial trends, respectively, as labelled.
Different line types correspond to different methods to estimate the
abundance ratios (see labels on the upper--right corner of the bottom
panel). Error bars are quoted at the $1\sigma$ level.
}
\label{fig:abund}
\end{figure}

In summary, our analysis shows that the spectra of XSG1 feature
homogeneously old stellar populations inside the effective radius,
with no significant age gradient. The absolute value of the age
depends on the methodology, a well-known issue plaguing
most stellar population studies (see, e.g.,~\citealt{Vazd:2001, Schiavon:2002}).
Throughout this work, we adopt as a reference the age estimates from
spectral fitting (method $FIT1$ above), rescaled to match the \hgs
--based value ($\sim 11$\,Gyr) at the galaxy centre. We stress that
this assumption serves only for illustrative purposes (e.g. to
estimate the metallicity and abundance ratio gradients of XSG1), while
in practice, when constraining the IMF, our fitting scheme takes into
account the uncertainties on the age zero-point with an ad-hoc
procedure (see Sec.~\ref{sec:method}).

The bottom panel of Fig.~\ref{fig:ageZ} shows the metallicity profile
for our reference age estimates, from the \mgfep\ index. The
\mgfep\ is largely independent of IMF slope, implying that, {\it at
  fixed age}, metallicity estimates are independent of the IMF (see
dotted and dashed black lines in the Figure). Notice that \mgfep\ is
also well-known for being independent of \afe~\citep{Thomas:03, TMJ11,
  Vazdekis:15}. Moreover, CvD12 models predict a small variation of
\mgfep\ with non-alpha elements (e.g. $\delta$\mgfep$\sim
-0.03$\,\AA\ for $\delta$\cfe$=+0.15$\,dex~\footnote{It is important
  to remind the reader that these predictions hold at fixed [Fe/H]=0,
  and assume no effect from varying abundance ratios on the
  solar-scaled isochrones.}, at $\sigma=300$\,\kms and for an age of
$13.5$\,Gyr). Hence, after accounting for the uncertainty on the
zero-point of the age, the \zh\ can be robustly constrained from
\mgfep . Fig.~\ref{fig:ageZ} shows that XSG1 has a negative
metallicity gradient of about $-0.25$\,dex per radial decade, a
typical value in massive early-type galaxies
(e.g.~\citealt{SanchezBlazquez:2006, Spolaor:2010}).

\subsection{Radial profiles of abundance ratios}
\label{sec:profs_xfe}

Deviations from the solar scale of the abundance ratios involving
single elements can partly mimic (or mask out) the effect of a
varying IMF on gravity-sensitive features
(see~\citealt{CvD12a}). Since TiO features, which are the main focus
of the present study, are mainly affected by \cfe\ and \afe\ (see
Sec.~\ref{sec:tio_profs}), in the present section we analyze
the radial gradients of \cfe\ and \afe\ in XSG1, pinpointing the
uncertainties arising from different model assumptions. Notice that we
do not aim at obtaining a precise, quantitative, determination of
abundance ratios, but rather to perform a qualitative characterization
of the radial behaviour of \afe\ and \cfe\ in XSG1.

Fig.~\ref{fig:abund} plots different estimates of \afe\ and \cfe\ (i.e. different methods/models; see
lines with different colours) as a function of galacto-centric
distance. The solid red-through-blue line, joining filled circles,
shows our solar-scaled proxy for \afe , measured as a difference
between the metallicity estimates derived, at fixed age, from
\mgb\ and \fet . The proxy has been calibrated onto
\afe\ with the aid of~\citet[hereafter TMJ11]{TMJ11} stellar
population models, as detailed in LB13 (see also VAZ15), resulting into an accuracy (rms) of 
only 0.025~dex in \afe . The dotted and
dashed lines in Fig.~\ref{fig:abund} also show \afe\ and \cfe\ obtained by
fitting simultaneously \mgfep , \mgb , \fet , and \cfs\ line-strengths
with predictions from MILES models with a Kroupa-like (\gammab=1.3)
and a bottom-heavy (\gammab=3) IMF, respectively. The fit is
done by minimizing the RMS scatter of model and observed line-strengths 
with respect to \cfe\ and \afe\ (see Eq.~\ref{eq:method} in Sec.~\ref{sec:method}),
where the sensitivity of line-strengths to \afe\ (\cfe ) is computed with
\amiles\ (CvD12) stellar population models, for old (13.5\,Gyr) age and
solar metallicity. The reason for adopting \mgb\ and \fet\ is that
these features are strongly sensitive to \mgfe , while \cfs\ is
extremely sensitive to carbon abundance
(e.g.~\citealt{TripiccoBell:1995}). The pink dotted lines in 
Fig.~\ref{fig:abund} are the same as the red-through-blue dotted lines, but estimating the
sensitivity of \mgb , \fet , and \cfs\ to \afe\ in the fitting with \amiles\ SSPs 
having super-solar (\zh$=+0.26$), rather than solar (see above) metallicity. In this case,
\cfs\  decreases more with \afe\ than for
\zh$=0$ , implying higher inferred values of \cfe . Notice also that,
as mentioned above, CvD12 models are computed at fixed [Fe/H],
rather than total metallicity. Hence combining CvD12 and extended MILES
model predictions is only meaningful for elements that give a
negligible contribution to \zh . Since this is only approximately true
for \cfe , we have also repeated the fits by modelling the sensitivity
of \cfs\ to carbon with TMJ11 models (computed at fixed \zh ). As
shown by the dot-dashed line in Fig.~\ref{fig:abund}, the TMJ11 models, relative
to CvD12, lead to lower \cfe\ estimates, by $\sim0.05$\,dex
(compare the dot-dashed versus dashed curves in Fig.~\ref{fig:abund}).

Fig.~\ref{fig:abund} shows that there is a good agreement among
estimates of \afe\ from different methods, with differences below
$\sim 0.08$\,dex for each radial bin. In particular, changing
\gammab\ (dotted vs.  dashed red-through-blue curves) gives
differences $\simlt 0.05$\,dex, implying that the derivation of
\afe\ is robust with respect to the assumed IMF slope. We find that
the radial profile of \afe\ in XSG1 is essentially constant with
radius, out to one half of the effective radius, with a very high
\afe\ ($\sim +0.4$), and decreases to $\sim 0.2$, 
only in the outermost bin. Regarding \cfe , differences among methods
are more significant (up to $0.1$\,dex) than for \afe . In general, we
find that, similar to \afe , \cfe\ is approximately constant with
radius ($\sim 0.2$\,dex), at least in the radial range from $0.2$ to
$0.5$\,\re , dropping to $\sim 0.1$\,dex at the largest distance
probed ($\sim 0.8$\,\re). Notice that although the decrease of \cfe\ in the innermost
radial bin might be due to model uncertainties in the high metallicity
regime, in practice such trend does not affect at all our IMF constraints (Sec.~\ref{sec:gamma_trends}). 
Also one can notice that, with the exception of the innermost bin, the ratio of \afe\ to
\cfe\ throughout the galaxy is approximately constant ($\sim 2$). The
radial behaviour of \afe\ and \cfe\ have important implications for
the analysis of TiO-based gravity-sensitive features, as discussed
below, in Sec.~\ref{sec:tio_profs}.

\subsection{Constraining the IMF slope}
\label{sec:method}
\subsubsection{Fitting scheme}
The approach we adopt to constrain the IMF is an improved version of
the method developed in our previous works (e.g.~FER13,
LB13,~\citealt{LB:15, MN:2015a, MN:2015b, MN:2015c}).  For each radial
bin of XSG1, we minimize the following expression,
\begin{eqnarray}
\rm \chi^2(IMF, t_M, {\rm [M/H]}, [X/Fe]_r )= \left[
  \frac{Age-Age_M}{\sigma_{Age}} \right]^2+ \nonumber \\
  +\rm \sum_i \left[ \frac{E_{O,i}^{ss}- 
  E_{M,i} }{\sigma_{E_{O,i}^{ss}}} \right]^2,
\label{eq:method}
\end{eqnarray} 
where the summation index $i$ runs over a selected set of spectral
features (see Sec.~\ref{sec:features}), $\rm Age$ and $\sigma_{\rm Age}$ are our reference estimates of
the age (updated through an iterative procedure; see Sec.~\ref{sec:corrs}) and its
uncertainty for the given spectrum (Sec.~\ref{sec:age_z}), $\rm
E_{M,i}$ are line-strength predictions for an SSP model with age $\rm
Age_M$, $\rm E_{O,i}^{ss}$ are observed line-strengths -- $\rm E_{O,i}$ --
corrected to solar-scale with an approach similar to that of LB13,
i.e.
\begin{equation}
 \rm E_{O,i}^{ss}=\rm E_{O_i}-C_{i}(Z_{MW}) \cdot [\alpha/{\rm Fe}],
\label{eq:corr}
 \end{equation}
where \ciz\ is the {\it observed} slope of the correlation between a
given index line-strength and \afe , derived from SDSS data (see
LB13).
$\rm Z_{MW}$ represents the total metallicity, estimated for a
Milky-Way-like IMF (in practice, we use a bimodal IMF with
\gammab=1.3) from the (emission-corrected)
\hbo\ vs. \mgfep\ diagram (see, e.g., Fig.~\ref{fig:hbo}). The
uncertainty, $\rm \sigma_{E_{O,i}^{ss}}$, is obtained by adding in
quadrature the statistical error on $E_{O,i}$ with the uncertainty on
the solar-scaled correction (see sec.~6 of LB13).

\begin{figure*}
\begin{center}
\leavevmode
\includegraphics[width=15.5cm]{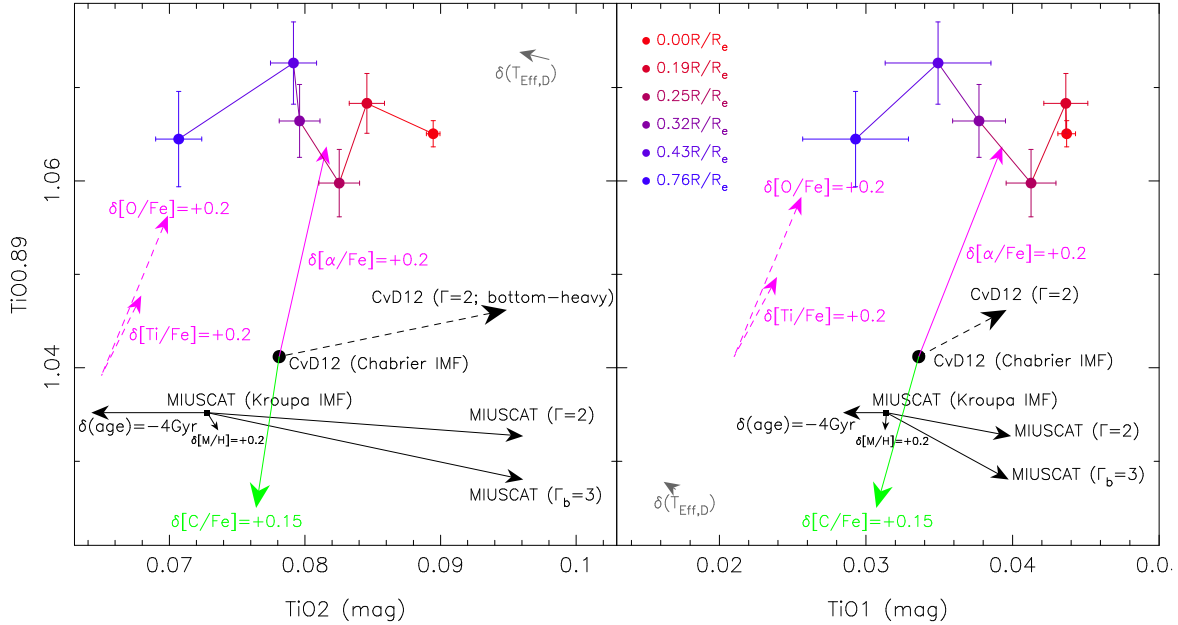}
\end{center}
\caption{Radial trend of the NIR TiO feature, \tionir , as a function
  of \tioii\ ({\sl left}) and \tioi\ ({\sl right}).  Red-through-blue
  dots, with error bars, correspond to different radial bins, as
  labelled in the panel on the right. The error bars are quoted
  at the $1\sigma$ level, and include uncertainties on sky
  subtraction. Solid (dashed) black arrows show the effect of varying
  the IMF from a Kroupa (Chabrier) to a bottom-heavy IMF, for MIUSCAT
  (CvD12) SSP models, with an age of 14 (13.5)\,Gyr and solar
  metallicity. For both CvD12 and extended-MILES models, we consider bottom-heavy
  unimodal distributions, with $\Gamma=2$, while the case
  of a bimodal bottom-heavy IMF, with \gammab$=3$, is also shown for
  the MIUSCAT models.  We also show the effect of a change in age and
  metallicity on the MIUSCAT models, for a Kroupa-like IMF.  All TiO
  features have very little dependence on \zh , while \tioii\ (and to
  less extent \tioi ) decrease with decreasing age. The effect of
  increasing \afe\ (\ofe , \tife ) by $+0.2$\,dex, in the CvD12 models is
  shown by the magenta solid (dashed) arrows, while the green arrows
  correspond to a variation of $+0.15$\,dex in \cfe .  Notice that
  \tionir\ is independent of IMF, while it is significantly more
  sensitive to abundance ratios than \tioii\ and \tioi\, allowing us to
  to break the degeneracy between abundance ratios and IMF slope. The grey
  arrows show the effect of changing the temperature of 
  dwarf stars (\dtd ) in the MIUSCAT models (see text
  for details). All data and model line-strengths refer to a velocity dispersion 
  of 300\,\kms\ (see Sec.~\ref{sec:models}) .
  }
\label{fig:tio_nocorr}
\end{figure*}

{ Notice that our IMF constraints for XSG1 rely on the use of
  single SSP models to fit observed line-strengths
  (Eq.~\ref{eq:method}).  As discussed in Sec.~\ref{sec:age_z}, this
  assumption is well motivated from results of spectral fitting, as
  well as from the high values of \afe\ for XSG1, pointing to
  homogeneously old ages for this galaxy at all radii.  Moreover, as
  discussed in LB13, the use of complex star-formation histories (e.g.
  2SSP models) does not affect significantly the derived IMF slopes in
  ETGs with high velocity dispersion, while the effect is measurable,
  although mild, in ETGs with low velocity dispersion. We note that
  spectral fitting (see {\sl FIT1} in \S\S\ref{sec:age_z}) yields
  probability distribution functions at all radii with a negligible
  fraction in young ($<$8\,Gyr) components. In~\citealt{MN:2015a}, we
  explicitly show that the use of multiple SSP models does not change
  the radial profile of the IMF slope in the elliptical galaxy NGC\,4552 whose
velocity dispersion is comparable to that of XSG1.}

\subsubsection{Selected spectral features}
\label{sec:features}
The definition of the main spectral features used in the present work is summarized in
Tab.~\ref{tab:indices}.
We focus on optical (mostly
TiO-based) gravity-sensitive features, i.e. \tioi , \tioii , \atio , and
\mgf~\footnote{We notice that a very similar index as \mgf , named
  $\rm bTiO$, has been defined by ~\citet{SPI:14} to constrain the
  IMF. In practice, $\rm bTiO$ and \mgf\ have a very similar sensitivity
  to IMF, and abundance ratios.}, contrasting the constraints from
these features with those from the Wing-Ford band (hereafter \feh ) in
the Near-Infrared ($\lambda\sim 0.992$\,$\mu$m).  Moreover, we
include the total metallicity indicator, \mgfep , in our $\chi^2$
minimization procedure.  Since \mgfep\ is independent of abundance
ratios (see, e.g., TMJ11, VAZ15), for this index we set \ciz$=0$ in
Eq.~\ref{eq:method}. Hence, the input of our fitting procedure
consists of empirically-corrected values of \tioi , \tioii ,
\atio , \mgf, plus \mgfep\ and our reference $Age$ estimates. Given
this input, we minimize Eq.~\ref{eq:method} over a set of extended MILES
models, deriving the best-fitting values of the relevant model
parameters, and in particular the IMF slope, i.e. $\Gamma$ (\gammab )
for unimodal (bimodal) models.

Notice that for \feh, no empirical correction is available at the moment, and thus the
fitting is performed in a different manner, relying on theoretical
(CvD12) models alone to estimate the impact of abundance ratios (see
Sec.~\ref{sec:feh} for details).  In addition to \tioi , \tioii , \atio ,
\mgf , and \feh , we also study the radial behaviour of the NIR TiO
feature, \tionir\ ($\lambda\sim 0.886$\,$\mu$m), defined by
CvD12. This feature is mostly sensitive to abundance ratios as well as
giant (rather than both dwarf and giant) stars in the IMF (in contrast to most TiO lines), and has been recently used, 
for similar purposes as in our work (i.e. studying IMF and
abundance-ratio gradients), by \citet{McC:2015}. These authors
combined \tionir\ with gravity-sensitive features arising from
different elements (Na and Fe) than TiO. To our knowledge, the present
work is the first one where optical and NIR TiO features -- sensitive
to the {\it same} elemental abundances (\tife\ and \ofe ; see
Sec.~\ref{sec:tio_profs}) are studied simultaneously to constrain the
IMF. Combining \tionir\ with features sensitive to temperature -- and
in particular the optical TiO-related indices -- allows us to
disentangle the contribution of giants and abundance ratios, with
respect to dwarfs, in the line-strengths. Notice that while we discuss
the radial behaviour of \tionir , we do not include it into our
$\chi^2$ minimization procedure (Eq.~\ref{eq:method}), as we fit IMF slope, age, and metallicity, 
but not abundance ratios, to the (solar-scale corrected) line-strengths.  
Indeed, obtaining a precise determination of abundance ratios is not the
main purpose of the present work.

\subsubsection{Novelties in the approach}
\label{sec:corrs}
The approach adopted in the present work (Eq.~\ref{eq:method}) has 
some relevant difference with respect to 
the one adopted in our previous works, i.e.
\begin{description}
 \item[1 - ] We do not consider any residual abundance-ratio term in
   Eq.~\ref{eq:method}, i.e. we assume that our solar-scaled
   correction is able to completely remove (within uncertainties) the
   combined effect of deviations of single elements from the solar
   pattern. This choice is motivated by the fact that corrected
   line-strengths are reasonably well fitted, at all radial bins, with
   solar-scaled models (see Sec.~\ref{sec:res});
 \item[2 - ] In order to account for
   uncertainties in the absolute calibration of the age (see
   Sec.~\ref{sec:age_z}), we apply an iterative procedure. First we
   minimize Eq.~\ref{eq:method} using our reference age estimate for each
   radial bin (see Sec.~\ref{sec:age_z}), then we re-scale {\it all} age
   estimates (i.e. for {\it all} bins) to the median value of the
   best-fitting $Age_M$ values, and repeat the fitting process. The procedure
   typically converges after two iterations, with differences between the
   average $\rm Age$ and $\rm Age_M$ less than a few percent.
 \item[3 - ] To apply the solar-scaled correction, $C_{i}(Z_{MW}) \cdot
   [\alpha/{\rm Fe}]$, we assume that the slope $C_{i}$ is a function
   of metallicity ($\rm Z_{MW}$), rather than velocity dispersion (as in
   LB13). This choice deserves some clarification.  In LB13, we binned
   SDSS spectra of ETGs with respect to central velocity dispersion, $\sigma$, and
   \afe~\footnote{In practice, we used the solar-scaled proxy for \afe,
     see above} , resulting into 18 stacks with $100<\sigma<300$~\kms,
   and a set of seven \afe\ sub-stacks at fixed $\sigma$. We showed
   that \afe\ provides a good estimate of the deviation of
   gravity-sensitive features from solar-scaled, in the sense that,
   after removing the {\it observed} departure of different indices
   from \afe$=0$ -- by means of the index--\afe\ correlations at fixed
   $\sigma$ -- one can fit a variety of features simultaneously with
   solar-scaled models.  In this regard, \afe\ should be seen as a
   metric that measures the deviation of a given spectrum from solar
   scale, and not only as an estimate of the actual enhancement of
   $\alpha$-elements.  Because of the \zh -- $\sigma$ relation in
   ETGs, metallicity also increases with $\sigma$ in the LB13
   stacks, and thus, the LB13 correction can be applied, in general,
   either as a function of metallicity, or $\sigma$. Based on the
   radial behaviour of gravity-sensitive features analyzed in this
   work (and in particular the \mgf\ indicator, see
   App.~\ref{app:atio_mgf}), we argue that a metallicity-driven
   correction provides the best way to generalize the LB13 approach
   also to the case of radially binned spectra (where the effect of
   $\sigma$ and \zh , are not equivalent, in contrast to SDSS
   data). In practice, for a given spectrum, we derive the
   metallicity from the \hbo\ vs. \mgfep\ diagram, using MIUSCAT SSP
   models with a Kroupa-like IMF (\gammab=1.3). We calculate the
   corresponding \ciz\ by interpolating the index-\afe\ slopes -- from the
   LB13 SDSS stacks -- with respect to $\rm Z_{MW}$ (estimated in the same
   way for the given spectrum and the SDSS stacks). We notice that our approach, 
   where the $C_{i}$ are a function of $\rm M_{MW}$, is also motivated by the 
   correlation of IMF slope and metallicity, recently found by~\citet[hereafter MN15b]{MN:2015b}.
   \end{description}
In a forthcoming paper (La Barbera et al., in preparation), we will
publish the coefficients \ciz\ for all gravity-sensitive features
covered by SDSS spectra, providing a detailed description of how one
should apply our empirical corrections to galaxy spectra.

\section{Results}
\label{sec:res}

We start by discussing the radial profiles of IMF-sensitive features
in XSG1.  In Sec.~\ref{sec:tio_profs}, we focus on the \tioi , \tioii
, and \tionir\ spectral features, as they allow us to disentangle the
IMF trends with other population features, such as abundance ratios. The radial behaviour of other
IMF-sensitive optical features (i.e. \atio\ and \mgf ) is discussed in
App.~\ref{app:atio_mgf}.  In Sec.~\ref{sec:feh}, we perform a
qualitative comparison of observations and model predictions for the
Wing-Ford band.  Sec.~\ref{sec:gamma_trends} shows results of our
fitting procedure applied to all observed line-strengths, i.e. the
relation between IMF-slope and galacto-centric distance in XSG1;
Sec.~\ref{sec:alpha_trends} deals with implications for the stellar
mass-to-light ratio gradient in XSG1, in comparison with dynamical
constraints.

\subsection{Radial profiles of optical and NIR TiO features}
\label{sec:tio_profs}

\subsubsection{Disentangling IMF from other effects}
Fig.~\ref{fig:tio_nocorr} shows one of the main results presented here,
namely the radial trends in the \tionir\ vs.  \tioii\ ({\sl left}) and
\tionir\ vs.  \tioi\ diagrams ({\sl right}) for XSG1. { In order to perform a qualitative analysis of these trends, the Figure also shows predictions of different stellar population models.}
According to CvD12 models, all three features are sensitive to \cfe\ and \afe\ (see
pink and green solid arrows in the Figure), in such a way that the
effect of an enhancement in \cfe\ tends to cancel out an increase in
\afe . The sensitivity to \afe\ mostly stems from \tife\ and \ofe , as
shown by the dashed pink arrows in the Figure. However, \tionir\ is
far more sensitive to abundance ratios than \tioi\ and \tioii , while
it is insensitive to the IMF. In fact, changes of the IMF slope
(either unimodal or bimodal) correspond to an almost horizontal shift
in both diagrams, as shown by the black solid (MIUSCAT) and dashed
(CvD12) lines in the Figure. In other words, the effect of abundance
ratios and IMF are orthogonal, and thus can be singled out in the
plots of \tionir\ vs. both \tioii\ and \tioi . The observed
line-strengths for XSG1 (plotted with red through blue circles, moving
from the innermost to the outermost bin) show a radial gradient {\it
  only for \tioii\ and \tioi }, while \tionir\ is almost constant out
to the largest radius probed. Notice that the lack of gradient for
\tionir\ is consistent with the competing effect between \afe\ and
\cfe\, and the fact that the ratio of \afe\ to \cfe\ is roughly
constant with radius in XSG1 (see Sec.~\ref{sec:profs_xfe}).  The grey
arrows in Fig.~\ref{fig:tio_nocorr} also show the effect of varying
the effective temperature
of dwarfs~\footnote{
Decreasing the temperature of giants, $T_{\rm Eff,G}$, can also (partly) mimic a varying IMF
in the \tioi\ and \tioii\ indices (see MN15a). However, as shown by~\citet{SPI:15}, when taking 
total metallicity into account, the effect of $T_{\rm Eff,G}$ is irrelevant to 
constrain the IMF. Varying $T_{\rm Eff,G}$ produces an ``oblique'', rather than horizontal (IMF), 
shift in the \tionir\ vs. \tioii /\tioi\ plots, in contrast to the observed (horizontal) trends. 
}, by adopting the prescription from~\citet{Pols:95}, cooler
than the MIUSCAT reference temperature for dwarf stars~\citep[see][]{Vazdekis:12}. 
Varying $\rm T_{Eff,D}$
has a minor effect on both \tioi\ and
\tioii , and thus cannot explain the observed gradients.
Regarding the sensitivity of TiO features to other stellar population
properties, we notice that, for \zh$>-0.4$, MIUSCAT models predict
both \tioi\ and \tioii\ to be independent of (total) metallicity (see,
e.g.,~\citealt{MN:2015b}). For old ages, \tioii\ (and to less
extent \tioi ) tend to decrease with decreasing age, in such a way
that a change in age by $\sim 4$\,Gyr, from an old centre to younger
outskirts (solid arrows, labelled $\delta(age)=-4$\,Gyr) 
might account for $\sim1/2$ ($1/5$) of the observed radial
gradient of \tioii\ (\tioi ). In practice, however, XSG1 has no
significant radial gradient in age (Sec.~\ref{sec:age_z}), and
thus a varying IMF is left as the only possible explanation for the
\tionir\ vs. \tioii\ and \tioi\ trends.

As mentioned above (Sec.~\ref{sec:method}), the use of \tionir\ in
combination with a dwarf-sensitive indicator (e.g. the Wing-Ford band)
has been originally proposed by CvD12, to disentangle the effect of
giants and dwarfs on the IMF.  The present work is the first attempt
to combine \tionir\ and the optical TiO features (in particular, \tioii\ and
\tioi ) to disentangle the effect of the IMF with respect to giant stars {\it
  and} abundance ratios. Recently, \citet{McC:2015} have used \tionir\ to
investigate the origin of radial trends of IMF-sensitive features in
two nearby ETGs, with $\sigma=230$\,\kms\ (NGC$\,$1023) and
$\sigma=245$\,\kms\ (NGC$\,$2974). They found a
slightly-decreasing radial trend of \tionir\ for NGC$\,$1023 (with a
variation of $\sim 0.01$ out to 1\,\re), and a constant trend for
NGC$\,$2974, the latter being consistent with our findings for XSG1.

\begin{figure*}
\begin{center}
\leavevmode
\includegraphics[width=15.5cm]{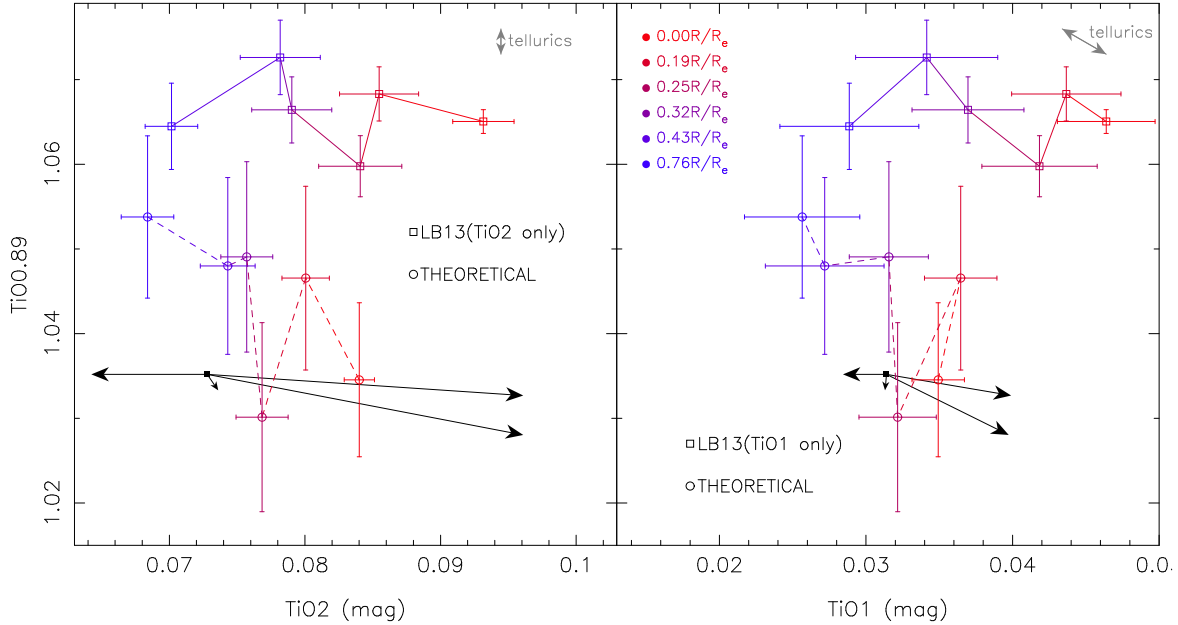}
\end{center}
\caption{ Radial trends of TiO features, as in
  Fig.~\ref{fig:tio_nocorr}, but after subtracting off the expected
  effect of the estimated \afe\ and \cfe\ abundance ratios from each
  spectral index (only from TiO2 and TiO1 in the LB13 approach, see the labels). 
  Error bars account for the uncertainties on the
  \afe\ and \cfe\ corrections from different methods (see 
  text). Notice the fairly good matching of the {\it theoretically} corrected EWs of \tionir\ with
  predictions from MIUSCAT models. The radial gradients of TiO indices
  are remarkably consistent with a radially varying IMF.
  Bimodal and unimodal models are fully degenerate
  in both diagrams, consistent with the results in LB13.
  Grey arrows show variations in the line-strengths of XSG1 from different methods used to 
  remove the telluric lines (see the text).
  Black arrows are the same MIUSCAT model predictions (with varying age, metallicity, and IMF) as in 
  Fig.~\ref{fig:tio_corr}.
  }
\label{fig:tio_corr}
\end{figure*}

\subsubsection{Abundance ratios: theoretical predictions}
While there is some difference between solar-scaled CvD12
and MIUSCAT model predictions for TiO features (see black circle and
small black square in Fig.~\ref{fig:tio_nocorr}), the most remarkable
offset in the Figure is the one between the \tionir\ model predictions
and the observed points.  The observed \tionir\ is significantly
higher (by $\sim$0.03) than in CvD12 and MIUSCAT models. This might be
driven by non-solar abundance ratios (i.e. \afe\ and \cfe ), or some
other effects~\footnote{
In particular, CvD12 showed that \tionir\ is strongly
sensitive to giant stars (see section~4.3 and figure~16 of CvD12).
As mentioned
above, the lack of a gradient in \tionir\ for XSG1 points to no 
significant radial change in the properties of
giants, but rather a genuine radial variation of the dwarf-to-giant ratio in
the IMF.  
}.  
 The effect of abundance ratios is further illustrated in Fig.~\ref{fig:tio_corr}, where we use CvD12 (theoretical)
model predictions to remove the expected effect of \afe\ and
\cfe\ from all TiO lines (see empty circles in the plot). In practice,
for each feature, we estimate the difference of line-strengths between
the \afe- (\cfe-) enhanced CvD12 SSP, and the solar-scaled one, for
old age ($13.5$\,Gyr), a Chabrier IMF, and the same velocity
dispersion as the observed spectrum at a given radial bin. We rescale
these differences to the measured values of \afe\ (\cfe) for XSG1. The
rescaled line-strength differences (i.e. the {\it theoretical
  abundance corrections}) are subtracted off from the observed values.
The error bars account for the statistical errors on line-strengths as
well as for uncertainties on \cfe\ and \afe\ from different methods
(Sec.~\ref{sec:profs_xfe}). Remarkably, the theoretically corrected values of
\tionir\ are fairly consistent with MIUSCAT (and CvD12)
solar-scaled predictions.
However, one should notice that theoretical corrections are
estimated at a specific point in parameter space (i.e. the CvD12 models
provide a grid of values for the age, IMF, and metallicity), and might be not exactly 
the same for . More
importantly, theoretical stellar population models rely on
ingredients, such as the stellar atmosphere calculations, that carry
notoriously difficult uncertainties.  Hence, we use theoretical
corrections in a qualitative manner throughout the present work,
whereas we rely on {\it empirical abundance corrections} (see
also Sec.~\ref{sec:method}) to perform a quantitative analysis.

\subsubsection{Abundance ratios: empirical corrections}
Fig.~\ref{fig:tio_corr} compares the theoretical (circles) and
empirical (squares) abundance corrections on \tioi\ and
\tioii\ line-strengths. Notice that for \tionir\ only the theoretical
approach is currently possible, as this feature is not covered by the
SDSS spectra analyzed in our previous works.  { Therefore, the
  offset between empirically-corrected points (squares) and model
  predictions (black arrows) for \tionir\ in Fig.~\ref{fig:tio_corr}
  should not be regarded as a discrepancy between models and
  data, but rather as the effect of abundance ratios that, at 
  present, cannot be described with the LB13 ``empirical''
  approach. Since \tionir\, is not used in the fitting
  procedure, the above offset does not affect any of the quantitative
  estimates regarding IMF variations in XSG1
  (Sec.~\ref{sec:gamma_trends}).}  In LB13, we found that \tioi\ and
\tioii\ do not depend much on the abundance ratio at fixed velocity
dispersion (metallicity).  More precisely, the TiO line-strengths are
found to increase (decrease) slightly with abundance ratio at the
lowest (highest) velocity dispersions (i.e.  metallicity, in our
refined approach, see Sec.~\ref{sec:method}, and section~5.2 of LB13).
Figs.~\ref{fig:tio_nocorr} and~\ref{fig:tio_corr} shows that this
small dependence on abundance ratio likely arises, {\it
  qualitatively}, from the counteracting effect of \afe\ and \cfe . As
a result, the empirical corrections tend to stretch out the radial
variations of \tioi\ and \tioii, making the radial gradient of
\tioi\ (and also \tioii ) even more pronounced than in the original,
uncorrected data.  We add in quadrature the statistical errors of
\tioi\ and \tioii\ to the error budget derived from our empirical
correction procedure, giving the error bars shown at the open squares
in Fig.~\ref{fig:tio_corr}.  Notice that although ideally theoretical
and empirical corrections should agree, in practice they do not.  The
theoretical approach tends to shift both \tioi\ and \tioii\ towards
lower values, in contrast to the empirical one, where the trends are
stretched out towards higher values.  This might be due to limitations
in the theoretical estimates (see above).  In addition, a combination
of departures of other element abundance ratios, such as \mgfe ,
\sife\ -- that individually do not produce a major change in the TiO
line strengths-- can compensate, along with \cfe, the increase in
\tioi\ and \tioii\ due to \ofe\ and \tife.  In practice, when
inferring the IMF, the fact that \tioi\ and \tioii\ are lower when
relying on theoretical abundance corrections would be compensated by a
younger age in our fitting procedure, implying similar IMF-radial
trends\footnote{In fact, since the sensitivity of the TiO features to
  the IMF slope is lower in younger populations, the use of either
  theoretical or empirical corrections would result in similar IMF
  gradients for XSG1.}  as those presented in
Sec.\ref{sec:gamma_trends}.  Since we have not been able to find a
consistent interpretation of all the observed indices when relying on
theoretical models (see below and App.~\ref{app:atio_mgf}) and because
of the current limitations of the publicly available CvD12 models (see
above), our quantitative analysis relies fully on the empirical
approach.

\subsubsection{Possible issues with data reduction}
One might wonder to what extent the trends in
Figs.~\ref{fig:tio_nocorr} and~\ref{fig:tio_corr} might be driven by
an issue related to the data reduction process. The grey arrows in
Fig.~\ref{fig:tio_corr} show the effect of a change in the telluric
correction method in the reduction process (i.e. telluric-standard
stars vs. {\sc molecfit}, see Sec.~\ref{sec:obs}) on TiO
line-strengths. The effect is negligible for \tionir\ and
\tioii , while it is mild for \tioi . Notice that although the spectral region
of \tionir\ is affected by telluric lines, the telluric correction is remarkably accurate (see also
Fig.~\ref{fig:spectra}), thanks to the excellent quality -- especially
the high resolution --  of our X-SHOOTER data.  The telluric correction
uncertainty is added in quadrature to the error bars on the
empirically corrected line-strengths. Other changes in the reduction
process (e.g. sky subtraction) have been found not to affect at all
the observed TiO line-strengths.

\begin{figure}
\begin{center}
\leavevmode
\includegraphics[width=8.7cm]{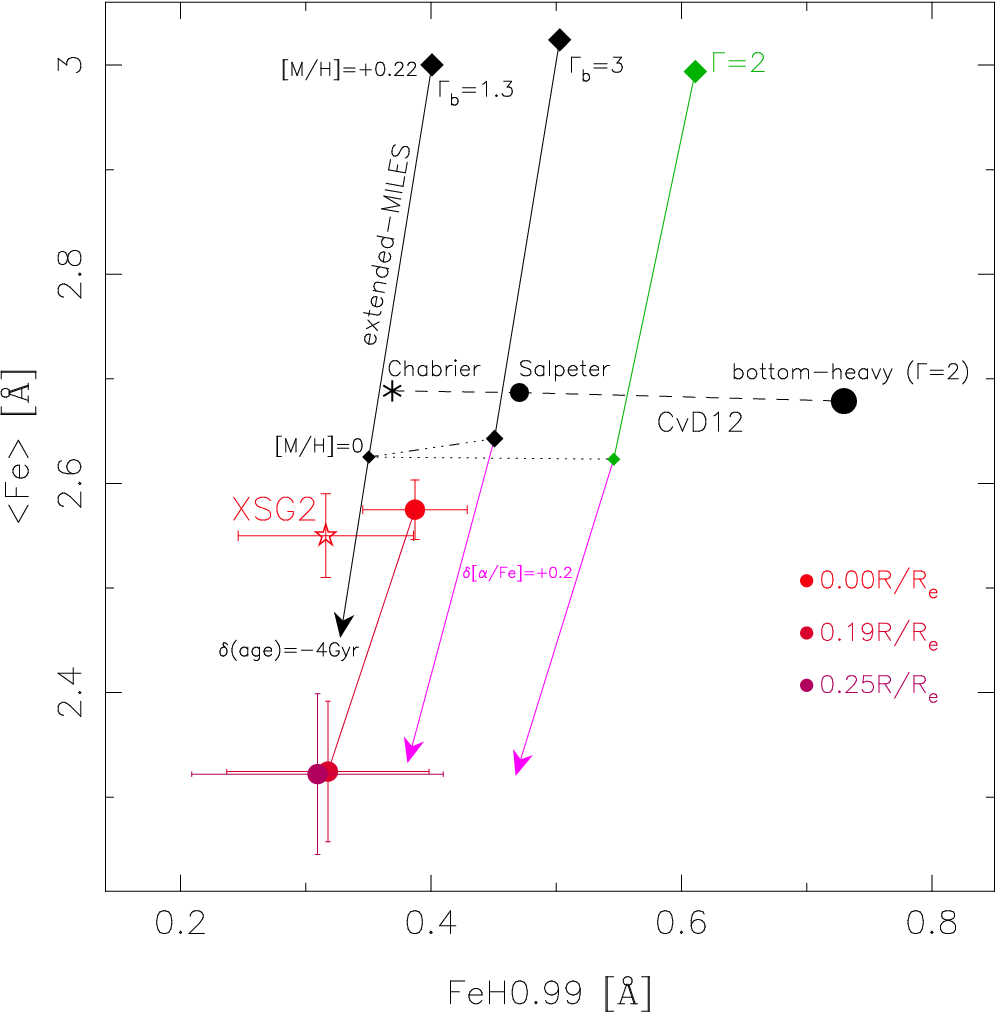}
\end{center}
\caption{Radial trend of the optical iron indicator, \fem , as a
  function of the Wing-Ford band, \feh .  Circles with different
  colours (from red to purple, as labelled) mark different radial
  bins for XSG1. The empty star corresponds to another massive ($\sigma
  \! \sim \!  300$\,\kms ; \afe$\sim 0.25$) ETG, XSG2 (see the text). Error bars mark
  $1\sigma$ uncertainties on line-strengths.  We show predictions for
  extended-MILES bimodal (solid-black lines) and unimodal (solid-green
  line) models with solar and super-solar \zh (plotted as small and
  big diamonds, respectively, for an age of $12.6$\,Gyr), as well as
  predictions for CvD12 unimodal models (dashed line), with a Chabrier,
  Salpeter and bottom-heavy ($\Gamma = 2$) IMF, at solar metallicity,
  and an age of $13.5$\,Gyr. 
   The dotted (dot-dashed) line joins predictions for unimodal (bimodal) extended-MILES models with 
  solar \zh\ (and should be compared with the dashed line for CvD12 models).
  Both \fem\ and \feh\ decrease with
  increasing \afe , as shown by the pink arrows, for extended-MILES models
  with \gammab$=3$ and $\Gamma =2$. Notice that the effect of
  \afe\ for \gammab$=1.3$ (not shown to make the plot more readable)
  is very similar to the case \gammab$=3$. All data and model
  line-strengths refer to a velocity dispersion of 300\,\kms\ (see Sec.~\ref{sec:models}) . }
\label{fig:feh_rad}
\end{figure}

\subsection{Constraints from the Wing-Ford band}
\label{sec:feh}

The optical gravity-sensitive spectral features constrain the mass
fraction of low-mass stars in the IMF (see LB13), and thus do not
allow for different functional forms of the IMF to be singled
out. This is illustrated in Fig.~\ref{fig:tio_nocorr}, where one can
see that an increase of the IMF slope of both bimodal and unimodal
models produces the same shift in \tioi\ and \tioii , and only a
tiny difference, with respect to the observed scatter, in \tionir . 
On the other hand, the Wing-Ford band, \feh , is sensitive to very
low-mass stars in the IMF (see figure~17 of CvD12). Hence, a
combination of this feature with optical indicators allows to better
constrain the functional form of the IMF at the low-mass end,
discriminating, for instance,  between unimodal and bimodal models.

Fig.~\ref{fig:feh_rad}, shows the abundance-sensitive ([Fe/H]) feature
\fem~$\rm =(Fe5270+Fe5335)/2$, in the optical, as a function of
\feh\ (see figure~12 of CvD12a). We show extended-MILES SSP predictions
for bimodal (solid-black lines) as well as unimodal (green line) models
with solar and super-solar metallicities (small and big diamonds,
respectively).  The extended-MILES predictions are compared to those for
CvD12 unimodal models (black-dashed line). 
Both sets of models agree for a MW-like IMF, for
which the \feh\ line-strengths from extended-MILES and CvD12 differ by only $\sim$0.03\,\AA .
  However, the change of \feh\ with $\Gamma$ (black-dotted for extended-MILES, 
  black-dashed line for CvD12 models) is
remarkably weaker in extended-MILES with respect to the CvD12 models.
Notice that for many IMF-sensitive indicators (e.g. Na- and
Ca-dependent features) the response to the IMF is also significantly
model-dependent
(e.g.~LB13, \citealt{SPI:15}). On the other hand, the behaviour of TiO features seems quite
robust between different models (see solid vs. dashed black lines in
Fig.~\ref{fig:tio_nocorr}).  Fig.~\ref{fig:feh_rad} shows that, using
the extended-MILES models, \feh\ increases with total metallicity
(CvD12a models do not vary \zh ), and increases with the IMF slope
more strongly for the unimodal case. The variation of the index is
much weaker for a bimodal IMF, where the low-mass end is tapered off
by construction.  The pink arrows in the Figure  show the effect
of varying \afe\ (by $+0.2$\,dex), at fixed \zh , on both \feh\ and
\fem. The \feh\ and \fem\ variation with \afe\ is computed from
$\alpha$-enhanced and solar-scaled CvD12 models (Chabrier IMF,
$13.5$\,Gyr, solar metallicity), removing the effect of the change in
total metallicity due to the fact that CvD12 models are computed at
fixed $\rm [Fe/H]$ (rather than \zh ). We remind the reader that, at
the moment, we do not have any empirical method to correct the
\feh\ for abundance variations (see Sec.~\ref{sec:tio_profs}). However, 
at least for \fem , we have verified that the predicted effect of \afe\ 
is robust, as the CvD12 model predictions are fairly consistent with 
those from the \amiles\ models. In fact, for $\rm \delta[\alpha/Fe]=+0.2$, 
we get $\delta$\fem$\sim\,-0.32$\,\AA\ from CvD12a models (see pink arrow in the
Figure for \gammab$=3$), while from \amiles\ we get $\delta$\fem\ in the range
from $-0.27$ to $-0.35$\,\AA, depending on age, metallicity, and IMF
(with the largest variations found in models at the highest \zh ).

Fig.~\ref{fig:feh_rad} also shows the observed \feh\ for XSG1 as a
function of galacto-centric distance, with red-through-purple filled
circles. Despite of the high S/N ratio of our X-SHOOTER data,
the estimates in the outermost radial bins have large uncertainties in
\feh\ to provide useful constraints on the IMF. Therefore, we focus
here on the three innermost radial bins. The most remarkable aspect in
the Figure is that the observed \feh\ is significantly lower than
expected for a bottom-heavy unimodal IMF -- compare the largest black
circle (CvD12 bottom-heavy estimate) with the red dot with error bars
(observed data).  This mismatch also applies to the extended-MILES
models. Specifically, for an old, metal-rich population
($\sim$12.6\,Gyr; \zh$\sim+0.25$~dex),  typical values in the central
regions of XSG1 (see Fig.~\ref{fig:ageZ}), and $\Gamma = 2$ (an IMF
slope even lower than the one implied by optical IMF-sensitive
indicators alone, see Fig.~\ref{fig:gamma_rad}), the extended-MILES
models give \feh $\sim 0.64$\,\AA\ ($\sigma=300$\,\kms).  For
\afe$\sim 0.4$ (see Fig.~\ref{fig:abund}), the \feh\ should be lower
by $\sim 0.15$\,\AA , implying an expected value of $\sim 0.49$\,\AA\ 
in the innermost radial bin, in disagreement with the measured value
of \feh$= 0.39 \pm 0.04$\,\AA .  Instead, for a bottom-heavy IMF
with \gammab$=3$ (12.6\,Gyr, \zh$\sim+0.25$), the extended-MILES prediction
(accounting for \afe ) is \feh$\sim 0.38$\,\AA, fully consistent with
the observations.  A similar kind of discrepancy with a unimodal IMF
is present in the second and third radial bins of XSG1, although it is
less significant because of the larger uncertainties on \feh .  We
emphasize that the sensitivity of \feh\ with respect to \afe\ is computed
in Fig.\ref{fig:feh_rad} by use of the CvD12 models for a Chabrier
IMF. If such sensitivity were larger for a unimodal bottom-heavy case, relative
to a Chabrier IMF, the \feh\ might decrease more strongly with
\afe\ than shown in Fig.~\ref{fig:feh_rad}, making this discrepancy
less significant. However, it seems not to be the case:  The second
massive galaxy, XSG2 (with $\sigma \sim 300$\,\kms ), observed as part
of the same programme (Sec.~\ref{sec:obs}), is shown as an empty
star with error bars in Fig.~\ref{fig:feh_rad}. XSG2 has a {\it  lower}
\afe ($ \sim 0.25$) than XSG1, and its \feh\ is far too-low with
respect to the bottom-heavy unimodal predictions. In the
following Section, we show that a bottom-heavy bimodal IMF  matches
both optical indicators and the Wing-Ford band in XSG1.

\begin{figure}
\begin{center}
\leavevmode
\includegraphics[width=8.5cm]{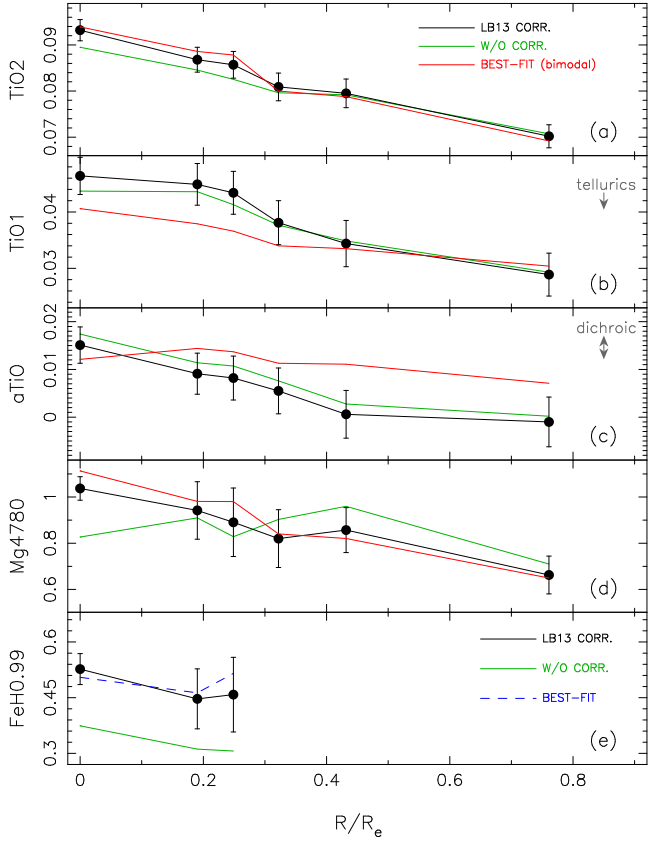}
\end{center}
\caption{Comparison of observed and best-fitting IMF-sensitive
  features as a function of galacto-centric distance, $\rm R/R_e$, in
  XSG1. In panels a, b, c, and d, the observed equivalent widths,
  empirically corrected to solar-scale (see Sec.~\ref{sec:method}),
  are plotted in black. The green curves are the uncorrected values,
  and the bimodal best-fit line-strengths, obtained by simultaneously fitting
  all the optical indicators (\tioi , \tioii , \atio , and \mgf ) are
  shown in red. In panel e (Wing-Ford band), the corrected
  values are obtained by removing the effect of the best-fitting \afe\ from
  the measured line-strengths (see
  Sec.~\ref{sec:feh}), while  the blue line shows extended-MILES (solar-scaled) 
  best-fitting line-strengths. Error bars denote $1\sigma$ uncertainties, including
  errors on the correction procedure, as well as the telluric correction
  (see grey arrow in panel b) and dichroic continuum variations (see
  grey arrows in panel c). All line-strengths have been corrected,
  based on the best-fitting extended-MILES model in each bin, to
  $\sigma=300$\,\kms\ (see Sec.~\ref{sec:models}).  }
\label{fig:ind_rad}
\end{figure}

\subsection{Radial trends of IMF slope}
\label{sec:gamma_trends}

By use of the method described in Sec.~\ref{sec:method}, we fit
IMF-sensitive features of XSG1 with either unimodal or bimodal extended-MILES
SSP models. Fig.~\ref{fig:ind_rad} shows the fitting results for bimodal models, plotting 
\tioi , \tioii, \atio , \mgf, and \feh\ as a function of galacto-centric distance. 
The unimodal fits are not shown in the Figure, as the comparison of best-fitting and observed line-strengths 
is very similar to that for bimodal SSPs.

We fit separately (i) the optical indicators and (ii) the Wing-Ford band:

\begin{description}
 \item[(i) ] For the optical indices (\tioi , \tioii, \atio , and \mgf), we have applied the 
 empirical corrections with respect to abundance pattern, and used
 our iterative approach to account for the uncertainty on the absolute age zeropoint, 
 as described in Sec.~\ref{sec:method}.
 \item[(ii) ] The fit with \feh\ uses simultaneously \feh, \mgb,
   \mgfep, and \fem\ without any iterative approach, but assuming the
   age constraint (i.e. the term $Age$ in Eq.~\ref{eq:method}) from
   the fit of optical features (i). Indeed, the age constraints do not
   affect significantly the results for \feh , as this feature is
   largely independent of age (see Fig.~\ref{fig:ind_rad}).  In the
   fit for \feh , we do not apply the empirical corrections for
   abundance pattern, as they are not available for this
   feature. Instead, we include \afe\ as a free-fitting parameter,
   using \feh, \mgb, \mgfep, and \fem\ to constrain, simultaneously,
   the parameters \zh , \afe , and the IMF slope.  To this effect, the
   sensitivity of \feh\ to \afe\ is modelled, at fixed \zh , as
   discussed in Sec.~\ref{sec:feh}.  We have found that our results
   (see below) remain unchanged if (a) we assume our proxy-based
   estimate of \afe\ in each radial bin (Fig.~\ref{fig:abund}), and
   fit simultaneously only \feh\ and \mgfep\ to infer \zh\ and IMF
   slope (either bimodal or unimodal), and (b) we also fix \zh\ to the
   output from the step (i), and use \feh\ only to infer the IMF
   slope.
 \end{description}

\begin{figure}
\begin{center}
\leavevmode
\includegraphics[width=8.6cm]{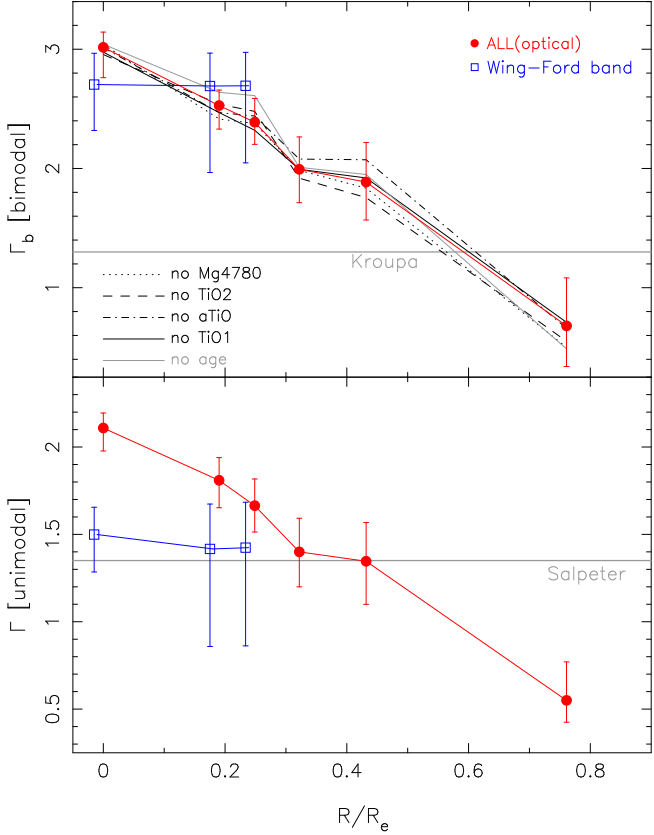}
\end{center}
\caption{Bimodal (top; \gammab ) and unimodal (bottom; $\Gamma$ ) IMF
  slope as a function of galacto-centric distance for XSG1.  Error
  bars denote $1\sigma$ uncertainties. Black lines and symbols
  correspond to different methods to constrain the slope, using
  different optical features, as labelled in the top panel.  Red
  curves and symbols show the IMF slopes derived by combining all the
  optical IMF-sensitive indicators analyzed in the present work
  (i.e. \tioi , \tioii , \atio , and \mgf ). Blue symbols and curves
  plot the IMF slope inferred using only the Wing-Ford band. Notice
  that in this latter case only the three innermost radial bins are
  plotted, as measurement errors on \feh\ do not allow us to obtain
  significant constraints for the three outermost bins.  For a bimodal
  IMF, we find an excellent agreement among optical indicators and
  \feh , while unimodal IMF slopes, inferred from the \feh , are on
  average too low (at more than the $3\sigma$ level) with respect to
  the optical constraints. { Notice that the inferred radial
profile of \gammab, is consistent when
excluding any one index from the analysis (see different line-types, in black), 
as well as when completely neglecting the age constraints in
the fitting procedure (grey solid curve; see the text)}.}
\label{fig:gamma_rad}
\end{figure}

Black symbols and lines in Fig.~\ref{fig:ind_rad} refer to corrected
indices. Uncorrected values are plotted as green lines. All
line-strengths in the Figure are corrected to a common velocity dispersion of
300\,\kms\ using the best-fit from the (bimodal) extended-MILES models (see
Sec.~\ref{sec:models}).
The best-fitting line-strengths are shown as red and blue lines in
Fig.~\ref{fig:ind_rad} for the optical and \feh\ indices, respectively. The 
best-fit values of $\Gamma$ (unimodal) and \gammab\ (bimodal) are
shown in the lower and upper panels of Fig.~\ref{fig:gamma_rad},
respectively. We also display, with different line-types, the results
for bimodal models with different combinations of optical indicators.
We also show the result when lifting the constraint on the age 
(grey curve; where the first term in Eq.~\ref{eq:method} is neglected).

Figs.~\ref{fig:ind_rad} and~\ref{fig:gamma_rad} show the main results
of the present work, as follows:
\begin{description}
 \item[1 - ] All the optical (TiO-based) IMF-sensitive features in
   XSG1 show a significant radial gradient, implying an IMF gradient
   in this massive galaxy, from bottom-heavy in the centre to MW-like
   at $\rm R/R_e\simgt 0.5$. Fitting all the optical features
   simultaneously (red lines in the Figures), we measure a radial
   decrease from \gammab$\sim 3$ ($\Gamma \sim 2.1$) in the centre to
   \gammab$\sim 0.6$ ($\Gamma \sim 0.5$) at $\rm R/R_e \sim
   0.8$. Notice that all TiO features analyzed in the present work are
   insensitive to IMF slope variations when the IMF is top-heavier
   than Kroupa (i.e.  for \gammab$\simlt 1.3$, or $\Gamma\simlt 1$; see
   figures~11 and~18 in LB13).  The fact that the lower error bars for the outermost bin 
   in Fig.~\ref{fig:gamma_rad} are relatively small (going, e.g., from 
   $\Gamma \sim 0.55$ to $\sim 0.4$ for unimodal models) is only 
   due to the lowest boundary value of IMF slope ($\Gamma$ and $\rm \Gamma_b=0.3$) for
   extended-MILES models.   Hence, our results in the outermost
   radial bin of XSG1 can be considered fully consistent with a
   Kroupa-like IMF.   
 \item[2 - ] For unimodal models, the Wing-Ford band provides
   inconsistent constraints (i.e. significantly lower $\Gamma$) with
   respect to those from the optical indicators (blue and red dots in
   lower panel of Fig.~\ref{fig:gamma_rad}, respectively). In the
   innermost radial bin, the $\Gamma$ derived with \feh\ is smaller
   than the optical-based one by $-29 \pm 8 \%$, i.e.  the discrepancy
   is significant at more than $3.5\,\sigma$.  In particular, this result
   seems not to be specific to XSG1, as also for XSG2 -- the second
   galaxy we have observed with X-SHOOTER -- the \feh\ is far too low
   to be consistent with a unimodal IMF (see Sec.~\ref{sec:feh}).
 \item[3 - ] For a low-mass tapered bimodal IMF, we get fully
   consistent results between optical lines and the \feh . In the
   innermost bin, the difference between \feh- and optical-derived
   \gammab\ is $-10 \pm 10 \%$.
\end{description}

Regarding the quality of the fits (i.e. Fig.~\ref{fig:ind_rad}), one
should notice that:
\begin{description}
 \item[i. ] The fit using optical features (red curves in the Figure)
   describes remarkably well the \tioii\ and \mgf\ observed trends
   (black circles and lines). We emphasize that the uncorrected
   line-strengths of \mgf\ do not show any clear radial gradient
   (green curve in panel d). It is our empirical abundance-pattern
   correction that reveals a radial gradient in this feature, making
   it fully consistent with \tioii . Notice also that, as discussed in
   App.~\ref{app:atio_mgf}, in the innermost bin, the empirical
   correction is the same when applying it either as a function of
   velocity dispersion or metallicity. However, in the former, the
   external corrected values of \mgf\ in panel d would be too high
   with respect to the best-fitting solution.  This is consistent with
   what we reported in MN15a, where the radial gradient of \mgf\ was
   found to be too small with respect to that of other
   features, when applying our empirical approach in terms of velocity
   dispersion. Our refined method, where abundance-pattern
   corrections are performed as a function of metallicity,
   provides consistent results.
 \item[ii. ] For \tioi , the three innermost data points appear too
   high (at the $2\,\sigma$ level) with respect to the best-fits. This
   discrepancy might be explained by uncertainties in our telluric
   correction procedure (see Sec.~\ref{sec:tio_profs}). Using {\sc
     molecfit} rather than telluric standard stars in the reduction
   process (see Sec.~\ref{sec:obs}), we get systematically lower \tioi
   , by $\sim 0.003$\,mag (see grey arrow in panel b of
   Fig.~\ref{fig:ind_rad}), making \tioi\ more consistent with our
   best-fitting solution at all radii.
 \item[iii. ] For \atio , the best-fitting solution shows a shallow
   radial gradient. These results from two competing effects: \atio\
   is expected (from extended-MILES models) to increase with the IMF. However, 
   while the index is independent of metallicity for a Kroupa-like IMF, it
   {\sl decreases} with \zh\ for a bottom-heavy IMF. Hence, in the inner
   regions, i.e. towards higher \zh\ and IMF slope, one would not
   expect large variations in the index. These competing trends,
   as well as the strong sensitivity of \atio\ to the continuum (see
   App.~\ref{app:atio_mgf}) make this index, in practice, a less
   useful IMF indicator than \tioi , \tioii , and also \mgf . The fact that
   the observed \atio\ shows a gradient, but the best-fitting
   solution does not, might be due to a combination of uncertainties
   in the empirical correction of this index, along with 
   issues with the absolute/relative calibration of \atio\ because of the
   instability of the X-SHOOTER dichroic response with time (see
   Sec.~\ref{sec:obs}, App.~\ref{app:atio_mgf}, and grey arrows in
   panel c of Fig.~\ref{fig:ind_rad}).
 \item[iv. ] For the Wing-Ford band, we obtain excellent agreement
   between the observed and best-fit line-strengths (blue and black
   lines in panel e of the Figure), although such a result is to be
   expected, considering that, in this case, the only IMF-sensitive
   feature used in the fitting procedure is \feh.  As mentioned
   above, the remarkable result is that one gets very consistent
   constraints, for a {\it bimodal} IMF, between the optical features
   and the Wing-Ford band.
  \end{description}

{ Finally, we want to emphasize that, although there are a few
  discrepancies between the observed and best-fit optical indicators
  in some radial bins (i.e.  \tioi\ in the three innermost, and
  \atio\ in the two outermost bins), Fig.~\ref{fig:gamma_rad} shows
  that the quality of the fit is good, within error bars. Moreover,
  {\it the inferred radial profile of the bimodal IMF slope, \gammab,
    is consistent when excluding any one index from the analysis} (see
  different line-types in the Figure), i.e.  if we include in the
  fitting, for each bin, only those indices for which models give a
  better matching to the data.  This proves that our results are
  robust, and not affected by possible mismatches between data and
  models. We also emphasize that {\it when completely neglecting the
    age constraints} in the fitting procedure we get very consistent
  results to our reference IMF profile (solid red curve) in
  Fig.~\ref{fig:gamma_rad}, implying that our findings are not
  affected by the uncertainty on the absolute-age zeropoint. }

\subsection{Constraints on the normalization of the IMF}
\label{sec:alpha_trends}

\begin{figure}
\begin{center}
\leavevmode
\includegraphics[width=8.5cm]{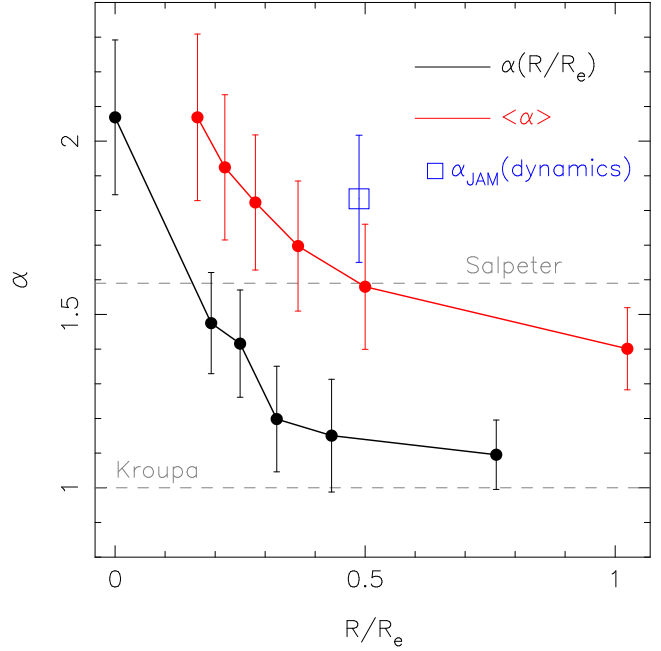}
\end{center}
\caption{Radial trend of the IMF mismatch parameter, $\alpha$, for
  bimodal IMF models (see text for details).  The black line and
  symbols show the local values of $\alpha$, while the luminosity-weighted
  value, $< \alpha >$, measured over circular apertures, is plotted in red. Error
  bars denote $1\sigma$ level uncertainties.  The blue square plots the mismatch 
  parameter, $\rm \alpha_{JAM}$, 
  obtained from the JAM dynamical analysis (Sec.~\ref{sec:kin}). 
  Notice that unimodal models would produce an overly high $< \alpha >$ ($\sim 3$)
  with respect to $\rm \alpha_{JAM}$.
  }
\label{fig:alpha}
\end{figure}

The radial trend of IMF slope found in XSG1 (Fig.~\ref{fig:gamma_rad})
leads to a radial gradient in the stellar mass-to-light ratio,
\mlstar.  To quantify it, we use the mismatch parameter $\rm
\alpha\equiv (M_*/L)/(M_*/L)_{MW}$, defined as the ratio between the
stellar \ml\ corresponding to the best-fit age, metallicity, and IMF
slope, and the one derived for the same age and metallicity, assuming
a MW-like IMF (we adopt as MW reference a bimodal \gammab$=1.3$
IMF). We derive $\alpha$ as a function of galacto-centric
distance. Notice that for a MW-like normalization, $\alpha \sim 1$,
whereas either bottom- or top-heavy IMF give $\alpha>1$ because of the
higher fraction of low-mass stars, or remnants, respectively. A
Salpeter normalization implies $\alpha \sim 1.6$.  At each radius, we
compute \mlstar\ from bimodal IMF models, as unimodal models are
inconsistent with constraints from optical features and the Wing-Ford
band.  To this effect, we random shift the values of \gammab , from
the best-fitting estimate at the given radius
(Fig.~\ref{fig:gamma_rad}), according to their uncertainties, and
compute the average \mlstar\ and its error.  The computation excludes
top-heavy models with \gammab$<0.5$, as these models have
\mlstar\ much larger than the MW reference (\gammab$\sim 1.3$), but
they are indistinguishable, concerning our IMF-sensitive indicators,
from the case with \gammab$ \sim 1.3$.  Fig.~\ref{fig:alpha} plots
$\alpha$ as a function of galacto-centric distance (black circles,
with $1\sigma$ error bars). As expected from the radial trend of
\gammab , the $\alpha$ parameter also decreases with $\rm R/R_e$, from
$\sim 2$ in the innermost radial bin, to $\sim 1$ beyond half of the
effective radius. The red curve in the Figure shows the integrated
constraints on $\alpha$, i.e. the luminosity-weighted values of
$\alpha$ computed within concentric circular apertures, under the
assumption of circular symmetry. As expected, the integrated version
of $\alpha$ shows a slower decline with radius, with respect to the
local one, becoming slightly lower than the Salpeter value already
within $R_e/2$ (second outermost red point, and dashed horizontal
line). Fig.~\ref{fig:alpha} also shows as a blue square
the mismatch parameter, $\rm \alpha_{JAM}$, obtained by normalizing the
total mass-to-light ratio from the Jeans Anisotropic Modelling
(Sec.~\ref{sec:kin}) to the stellar M/L for a Kroupa-like IMF (see
above).  Notice that we find $\rm (M/L)_{JAM}=7.5 \pm 0.75$, implying
$\rm \alpha_{JAM}=1.83 \sim 0.17$. $\rm (M/L)_{JAM}$ can be interpreted as the
ratio between total mass (including dark matter, which is likely
$\la 30\%$ within the central galaxy regions; see~\citealt{Cappellari:2013}) and the $r$-band luminosity
within a sphere of radius R$\sim$2''.  This is actually the region
where the data are more reliable and the fit using JAM describes them
better. For R$\sim$2'' ($\rm R/R_e \sim 0.5$), our bimodal-IMF fits
imply an aperture integrated value (see red circles in the Figure) of
$\rm \alpha \sim 1.57 $, i.e. smaller than $\rm \alpha_{JAM}$.  Hence, not only our
spectroscopic constraints are consistent between optical and NIR
indicators, but they are also consistent with the dynamical
analysis. The ratio $\rm(\alpha_{JAM}-\alpha)/\alpha_{JAM}$ gives the
dark-matter fraction (i.e. the ratio of non-stellar to total mass) for
XSG1, within half the effective radius, and amounts to $14 \pm
11\,\%$, consistent with the expectation that this galaxy should have
a minor dark-matter contribution in the central
region~\citep{Cappellari:15, Posacki:15}.

\section{Discussion}
\label{sec:disc}

\subsection{IMF gradients in ETGs}

The existence of a radial gradient of the IMF in XSG1
agrees well with our previous results for NGC\,4552 and NGC\,5557 --
two massive nearby ETGs with central velocity dispersion of $280$ and
$300$\,\kms , respectively (MN15a). For NGC\,4552 (the best
target in MN15a), the bimodal IMF slope decreases from
$\Gamma_{\rm b}\sim 3$ in the galaxy centre, to $\sim 2$ at $\rm R\sim 0.4R_e$,
consistent with the profile in Fig.~\ref{fig:gamma_rad} for
XSG1. Notice, however, that the velocity dispersion profile of XSG1
ranges from $\sim 340$\,\kms\ in the centre to $\sim 270$\,\kms\ at
$\rm R_e$.  Hence, our results extend the findings of MN15a at
significantly higher velocity dispersion, suggesting that IMF
gradients are common in high velocity dispersion ETGs. On the other
hand, we do not expect such gradients to be
universal. In~\citet[hereafter MN15c]{MN:2015c}, we derived the IMF
profile of the massive and compact ``relic'' galaxy NGC\,1277
($\sigma \sim 400$\,\kms), finding only a mild IMF gradient, with
\gammab\ decreasing from $\sim 3$ in the galaxy centre, to $\sim 2.5
\pm 0.4$ at $\rm R/R_e\sim 1$. In MN15b, analyzing a sample of ETGs
from the CALIFA survey (with $ 160\simlt\sigma\simlt 280$\,\kms ) we
found that locally -- i.e. at a given position within the galaxy --
\gammab\ is best correlated with total metallicity. 
Remarkably, no radial variation
is found in any of the observed properties of XSG1 (e.g. kinematics
profile, stellar population properties, abundance ratios), except for
IMF and metallicity, reinforcing the idea of a tight connection
between these two parameters. Similarly to NGC\,1277, XSG1 has a high
$\sigma$, and very high abundance ratios (\afe$\sim +0.4$;
comparable to NGC\,1277, with \afe$\sim +0.3$), with no age gradient
and no  radial variation of \afe\ within $\rm R_e/2$ (see
Fig.~\ref{fig:abund}). Nevertheless, XSG1 and NGC\,1277 have very
different radial variations of the IMF, suggesting that one may
expect, in general, a wide range of IMF profiles in massive ETGs. This
variety, also expected from the scatter of the IMF-\zh\ relation (see
MN15b), is similar to the observed metallicity gradients in ETGs. On
average, a correlation of \zh\ gradients with velocity dispersion is
detected (see, e.g.,~\citealt{SpiderIV}). However, this correlation is
tighter at low-velocity dispersion ($\sigma< 150$\,\kms), whereas a
wide scatter of metallicity gradients is observed in more massive ETGs
(see, e.g., figure~8 of~\citealt{Spolaor:2010}), ranging from almost
null to significantly negative ($\rm \delta[M/H]/\delta \log R \sim
-0.6$).  This scatter originates possibly from the different merging
histories of massive ETGs~\citep{Hopkins:2009}. Our results for XSG1
and other massive galaxies suggests that each specific channel of ETG
formation will shape the radial IMF profile of ETGs.  While an IMF
gradient will be produced by the different star-formation and chemical
enrichment processes at different radii during the initial stages --
likely through a time-varying IMF scenario~\citep{Weidner:2013,
  Ferreras:2015} -- this gradient is expected to be modified at later
times through the different merging events.  This scenario naturally
yields a range of IMF radial gradients in nearby ETGs, possibly
without a sharp correlation with any specific property of the galaxy.

While in our previous works (MN15a and MN15c) we presented evidence
against abundance ratios and other systematic effects, such as the
effective temperature of giants/dwarfs, as the major driver of radial
gradients of IMF-sensitive features in ETGs, we separate more
explicitly in XSG1 the effect of the IMF from other confounding
factors.  We find a significant radial gradient of features sensitive
to the dwarf-to-giant ratio in the IMF (e.g.~\tioi\ and~\tioii ),
whereas no gradient is detected either in the abundance ratios (\afe ;
\cfe ) or \tionir, sensitive to giants and abundance effects, but
insensitive to dwarfs). To our knowledge, this is the clearest
evidence so far, of a radial IMF gradient in a massive galaxy.  The
combination of no gradient in \afe\ and a radial trend in the IMF
confirms our previous claim that abundance ratio variations cannot be
the major driver of the observed trends (LB13 and~\citealt{LB:15}).  These
results are in good agreement with the relation between
dynamically-determined IMF mass normalization factors and stellar population
parameters~\citep{Smith:2014, McDermid:2014}.  Notice that a better
correlation of the IMF with \afe , rather than velocity dispersion,
has been originally suggested by~\citet{CvD12b}, but it is not
confirmed from studies where the effects of $\sigma$ and~\afe\ are
singled out. Indeed, the correlation of IMF with
\afe\ in~\citet{CvD12b} might be driven by the fact that CvD12 models
are computed at fixed [Fe/H], rather than total metallicity.

Recently,~\citet{McC:2015} investigated the origin of radial trends of
IMF-sensitive features in two nearby ETGs, with
$\sigma=230$\,\kms\ (NGC$\,$1023) and
$\sigma=245$\,\kms\ (NGC$\,$2974), respectively. They detected a
radial gradient in the dwarf-sensitive feature NaI8190
($\lambda\lambda \sim 8180-8200$\,\AA), with only a mild variation of
\feh , favouring abundance ratio variations (in particular, \nafe\ and
\nfe ) as the main explanation for the observed trends of
IMF-sensitive features in ETGs. Although the radial behaviour of the
IMF might be flat in some massive ETGs (see above), we emphasize that,
in practice, the presence of abundance- and IMF-gradients can 
coexist, hampering a direct interpretation of the indices.  One
example is the \mgf\ feature (bTiO; see \citealt{SPI:14}). No
significant -- at most a very shallow -- gradient is observed for this
IMF-sensitive feature for XSG1 as well as NGC\,4552 (see
MN15a). However, note that abundance ratios tend to wash out the
underlying IMF trend, and it is only by correcting for abundance
effects that one recovers consistent results with other features (see
App.~\ref{app:atio_mgf}). For some features, like \feh , the
interpretation is even trickier. According to predictions from new NIR
(extended MILES) stellar population models with varying total
metallicity {\it and} IMF, the \feh\ response to (total) metallicity
-- and thus also to \afe\ at fixed \zh -- is larger at higher
metallicity. Hence, even if \afe\ were constant (as in XSG1), the
\afe\ tends to further decrease the observed line-strength of \feh\ in
the galaxy centre with respect to the outer regions, counteracting the
effect of a steeper IMF in the centre (i.e. reducing the radial
gradient of \feh ).  Notice that in LB13 we also reported that the
abundance-ratio response of some gravity-sensitive features might be
coupled with the IMF slope. For instance, in LB13 (see figure~15) the
response of NaI8200 (CaT) to \nafe\ (\cafe ) was shown to be likely
stronger (weaker) for SSP models with a bottom-heavy IMF, relative to
Kroupa. This coupling, not included in any publicly-available version
of stellar population models with varying abundance ratios
(e.g. CvD12), might be responsible for some of the apparent
contradictions between model predictions and
observations~\citep{Smithetal:2015}.

\subsection{IMF shape and Mass-to-Light ratio gradients in ETGs}

Our analysis of optical and NIR features in XSG1 is inconsistent with
a unimodal, single power-law IMF, favouring a functional form tapered
off at the low-mass end, such as the bimodal IMF
of~\citet{Vazdekis:1996}.  We emphasize that this result does not
imply that the IMF functional form has to be bimodal.  As discussed in
FER13 and LB13, unimodal IMF models are also disfavoured by the fact
that they predict overly high stellar $\rm M/L$ ratios with respect to
simple dynamical constraints. However, a possible way out is to assume
a unimodal parametrization, increasing the low mass-end cutoff, $\rm
M_{low}$, in the IMF (with $\rm M_{low} \simlt 0.15\,M_\odot$;
see~\citealt{BARN:2013, SPI:15}). In fact, very-low mass stars ($\rm
<0.15 \, M_\odot$) do not contribute significantly to the integrated
galaxy light, while they provide a major contribution to the $\rm
M/L$. Since the Wing-Ford band is sensitive to the very low-mass stars
in the IMF (see figure~17 of CvD12), increasing $\rm M_{low}$ might
also reconcile optical and NIR indicators for XSG1, in the same way as
adopting a bimodal IMF. However, this option is rather contrived, and
would beg the question of why such a gap exists between low-mass stars
and massive brown dwarves. An alternative approach, taken
by~\citet{CvD12b}, is to assume a three-segment set of power laws for
the IMF, describing individually the behaviour of very-low, and low,
mass stars. In this parametrization, our results for XSG1 would imply
that the IMF has to be steep at low, but not at very-low mass. While
the information on the detailed shape of the IMF is imprinted in the
optical and NIR features of a galaxy spectrum (CvD12), it is not clear
if this information can be recovered, mostly because of
current uncertainties on stellar population models, and on 
modelling the effect of variations in the  abundance ratios
on the targeted spectral features. {\it Regardless
of the detailed shape of the IMF, one important result from the
present work is that spectroscopic constraints are inconsistent with a
pure unimodal parametrization of the IMF extended down to the H-core
burning stellar mass limit. } We notice that our dynamical analysis, based on JAM modelling with
an approach as close as possible to that of the ATLAS$^{\rm 3D}$ sample,
provides \ml\ constraints fully consistent with our stellar population
analysis.

Recently,~\citet[hereafter SLC15]{SLC:2015} analysed a sample of
nearby ETGs where \ml\ can be constrained through strong gravitational
lensing in the inner regions of galaxies. For three systems, SLC15
estimated a ``mass excess factor'', $\alpha$, only marginally
consistent with a Salpeter-IMF normalization ($\alpha \sim 1.6$), in
tension with stellar population models.  Among the three lensing
galaxies of SLC15, the more similar one to XSG1 is perhaps SNL-2, with
a velocity dispersion of $320 \pm 18$\,\kms , \afe$=+0.38 \pm 0.06$,
and an effective radius of $\sim 6$\,kpc (in the J band). Within the
Einstein radius ($\rm R_{Ein}$), that probes a region $\rm \sim
0.4\,R_e$ for SNL-2, SLC15 derive $\alpha=1.27 \pm 0.20$
($\alpha=0.94\pm0.17$), for the case where no dark matter (dark
matter) is assumed. At a galacto-centric distance of $\rm 0.4 \, R_e$,
XSG1 has an (aperture-)integrated value of $\alpha = 1.7 \pm 0.2$, for
a bimodal IMF.  This value is only marginally higher (below the
$1.5\sigma$ level) with respect to the no-dark-matter estimate of
SLC15. Moreover, as discussed above, one may expect, and we do
observe, a variety of IMF profiles in ETGs. Hence, we cannot exclude
that the IMF normalization is different when comparing different
galaxies (e.g. XSG1 and SNL-2). Furthermore, the estimate of the mass
excess, $\alpha$, changes when adopting different IMF
paramaterizations in the stellar population analysis (see FER13 and
LB13). The case of SNL-1 in SCL15 is also interesting. This lensing
galaxy has $356 \pm 18$\,\kms , \afe$=+0.31 \pm 0.05$, and an
effective radius of $\sim 2$\,kpc, being more compact than both XSG1,
and SNL-2. Within an $\rm R_{Ein}$ of $\rm 1.2 \, R_e$, SLC15 find
$\alpha= 1.42 \pm 0.15$ ($\alpha=1.20\pm 0.13$), for the case with
no-dark-matter (dark-matter).  In the case of XSG1, at the maximum
galacto-centric distance probed of $\rm \sim 1 \, R_e$, we find an
(aperture-)integrated mass excess $\alpha=1.40 \pm 0.12$ (see red
curve in Fig.~\ref{fig:alpha}), fully consistent with the estimate of
SLC15 for SNL-1. These comparisons show that radial IMF, and
mass-to-light, gradients are an important ingredient when comparing
constraints from different techniques.

\citet[hereafter SBK15]{SPI:15} have recently suggested an
anticorrelation between IMF slope and surface density, defined as $\rm
\rho = \sigma^2 / R_e^2$ in units of $\rm 1000\,(km s^{-1})^2/kpc^2$.
Since ETGs with high \afe\ are also the more compact ones around the
Fundamental Plane relation~\citep{Gargiulo:2009}, the result of SBK15
seems consistent with our findings (see~\citealt{LB:15}), namely, at fixed
velocity dispersion, the ETGs with the highest-\afe\ might also have
the lower IMF slopes.  However, our preliminary results for XSG2, the
second galaxy observed in our X-SHOOTER campaign, challenges this
scenario. XSG1 and XSG2 have $\rho \sim 5.5$ and $\sim 2$,
respectively, i.e. XSG2 is significantly less dense than
XSG1. Nevertheless, it has a lower \feh\ (see Fig.~\ref{fig:feh_rad}).
This result, combined with its higher central metallicity (\zh$\sim +0.35$)
and lower \afe\ ($\sim +0.25$) with respect to the values for XSG1
(\zh$\sim +0.25$ and \afe$\sim +0.4$), implies a lower IMF slope.
Moreover, \citet{Conroy:2013} found a mismatch parameter
 $\alpha = 2.27 \pm 0.16$ for stacked spectra of compact ETGs
in the SDSS, at $\sigma \sim 270$\,\kms , implying a higher IMF
normalization in more compact galaxies with respect to the general
population of ETGs (see, e.g., figure~5 of CvD12b).  We point out that
the presence of a variety of IMF gradients in massive ETGs 
complicates the comparison of IMF slopes among galaxies with different
degree of compactness. Larger samples are required to address this
issue in detail.

\section{Summary}
\label{sec:summary}

We present a detailed analysis of a set of targeted IMF-sensitive line
strengths in a massive ($\sigma\sim 300$\,\kms) early-type galaxy
(XSG1), at $z \sim 0.05$, observed with the VLT/X-SHOOTER
instrument. The lack of a significant gradient in the age and \afe --
two stellar population properties -- out to one half of the effective
radius (Figs.~\ref{fig:ageZ} and \ref{fig:abund}), makes this target
especially suited to the analysis of radial variations in the IMF. By
combining several TiO-based indices in the optical and NIR, we exploit
the different sensitivities to the underlying stellar population
parameters to disentangle the contribution from the stellar IMF. The
observed indices for XSG1 on the \tionir\ vs.  optical TiO diagrams
(Fig.~\ref{fig:tio_nocorr}) show a significant gradient in the optical
indices, not followed by the NIR index, which is insensitive to the
IMF.

Our findings strengthen the case for radial variations within an
effective radius of the IMF in massive ETGs, further supporting the
idea (MN15a) that the IMF is a local property of ETGs.  In XSG1, we
find a bottom-heavy population (\gammab$\sim$3) in the innermost
regions, gradually changing to a standard Kroupa-like IMF
(\gammab$=1.3$) at $\rm R\simgt R_e/2$ (Fig.~\ref{fig:gamma_rad}).
The lack of a measurable gradient in abundance ratios, over the
spatial scale where the IMF is found to vary, points to a possible
relationship between IMF and total metallicity in regions with high
velocity dispersion, in agreement with \citet{MN:2015b}.

In addition we include in the analysis the FeH-based Wing-Ford band,
sensitive to very-low mass stars. A remarkable mismatch is found when
adopting an IMF with a unimodal (single power law) functional
dependence. In contrast, a bimodal IMF (which substitutes the power
law at the low mass end with a constant value), is fully consistent
with all indicators (Fig.~\ref{fig:feh_rad}).  Hence, for the first
time, we are able to rule out the unimodal IMF based on a pure stellar
population analysis (i.e. beyond the fact that the unimodal stellar M/L is
overly high with respect to dynamical constraints; see LB13).

Remarkably, the dynamical analysis based on the Jeans anisotropic
modelling of \citet{Cappellari:2008} gives a very consistent stellar
M/L with respect to the (bimodal) model predictions of the
spectroscopic method via gravity-sensitive spectral features
(Fig.~\ref{fig:alpha}).

From a more technical point of view, the present work generalizes the
empirical approach of LB13 to deal with spectra at different
galacto-centric distances.  Remarkably, once {\it
  empirical solar-scale corrections} are applied as a function of
total metallicity (rather than velocity dispersion, see LB13, MN15a),
we are able to describe fairly well a variety of gravity-sensitive
features, including those for which a radial gradient is apparently
not detected (Mg4780).

In a future work, we plan to extend the present analysis, by combining
TiO- and \feh- based results with constraints from Ca and Na features,
to discuss how abundance ratios affect the combined use of these
features to constrain the stellar IMF.


\section*{Acknowledgments}
{
Based on observations made with ESO Telescopes at the Paranal
Observatory under programmes ID 092.B-0378 and 094.B-0747 (PI: FLB).
FLB acknowledges the Instituto de Astrof\'isica de Canarias for the
kind hospitality while this paper was in progress.  We thank dr.
J. Alcal\'a for the insightful discussions and help with the reduction
of XSHOOTER spectra.  We also thank the anonymous referee for his/her
useful comments that helped us to improve our manuscript.  We have
made extensive use of the SDSS database
(http://www.sdss.org/collaboration/credits.html).  MC acknowledges
support from a Royal Society University Research Fellowship.  We
acknowledge support from grant AYA2013-48226-C3-1-P from the Spanish
Ministry of Economy and Competitiveness (MINECO).}

\appendix

\section{Radial trends of the aTiO and Mg4780 IMF-sensitive features}
\label{app:atio_mgf}

In addition to the \tioi\ and \tioii\ indices, we also consider the
\atio\ and \mgf\ optical features to constrain the IMF in XSG1
(Sec.~\ref{sec:method}). In this appendix, we describe the radial
behaviour of these features. 

\subsection{The aTiO index}
\begin{figure}
\begin{center}
\leavevmode
\includegraphics[width=8.5cm]{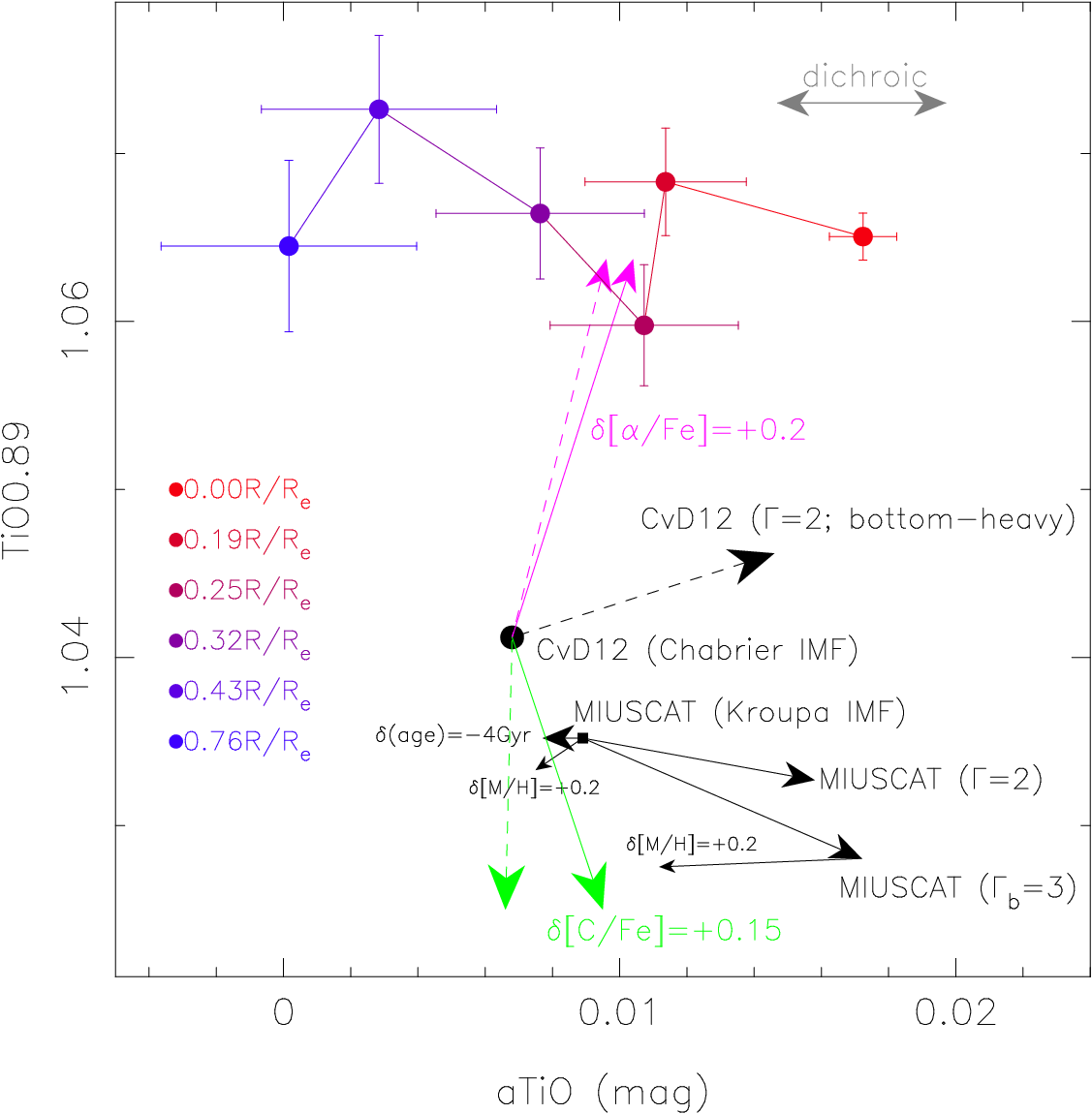}
\end{center}
\caption{Same as in Fig.~\ref{fig:tio_nocorr} but plotting \tionir\ as
  a function of the \atio\ IMF-sensitive feature. Pink and green
  dashed arrows show the sensitivity of \tionir\ and \atio\ to
  \afe\ and \cfe, respectively, when removing the continuum from CvD12
  models.
    The black two-sided
  arrow in the top-right of the plot shows the uncertainty in the
  absolute value of \atio\ because of the time-variable response of
  the dichroic of the X-SHOOTER spectrograph in the VIS arm.  }
\label{fig:atio_nocorr}
\end{figure}

\begin{figure}
\begin{center}
\leavevmode
\includegraphics[width=8.5cm]{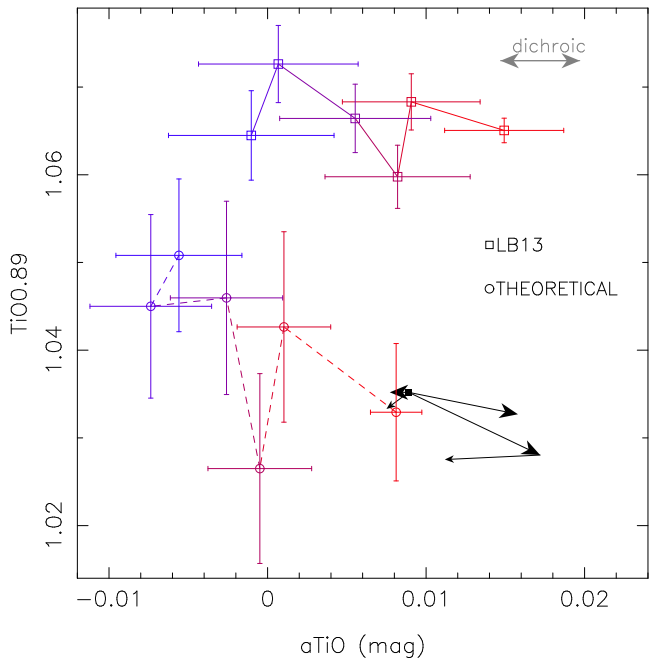}
\end{center}
\caption{Same as in Fig.~\ref{fig:atio_nocorr} but (i) correcting
  \atio\ and \tionir\ for their expected sensitivity to \afe\ and \cfe
  (see circles and dashed line), and (ii) correcting \atio\ for its
  SDSS-based sensitivity to abundance ratios, as in LB13 (see squares
  and solid line).  }
\label{fig:atio_corr}
\end{figure}

Fig.~\ref{fig:atio_nocorr} shows
\tionir\ (see Sec.~\ref{sec:tio_profs}), as a function of \atio , at
different galacto-centric distances (red through blue, from the centre
outwards). \atio\ increases with IMF slope, slightly more in
the bimodal case, in contrast to \tioi\ and \tioii, for
which the line-strengths of models with $\Gamma=2$ correspond exactly
to those for \gammab$=3$ (see Fig.~\ref{fig:tio_nocorr}). The index is
independent of age (see~\citealt{SPI:14}), while it is anticorrelated
with metallicity. As shown in Fig.~\ref{fig:atio_nocorr}, the
dependence on \zh\ is rather complex, as it is coupled with the IMF.
\atio\ decreases slightly with \zh\ for a MW-like IMF, while it
strongly decreases with \zh\ for a bottom-heavy distribution (see the
two black arrows in the Figure, for $\rm \delta[M/H]=+0.2$).  In other
words, \atio\ is, {\it in general}, not a good IMF-sensitive
indicator, as at high metallicity it is expected to have essentially
the same value for either a MW-like or a bottom-heavy IMF.  Hence, for
a galaxy with a bottom-heavier IMF in the centre, with respect
to the outer regions -- as it is the case for XSG1 -- one would expect
a shallow radial gradient in \atio .  In contrast,
Fig.~\ref{fig:atio_nocorr} shows a significant radial variation of
\atio\ in XSG1.  Using CvD12 models, we find that in addition to the
anticorrelation with $\rm[Fe/H]$ (consistent to the one with \zh\ in the
MIUSCAT models), \atio\ also depends on \afe\ and \cfe . As for
\tioi\ and \tioii , the index increases with \afe . However, this
dependence is not compensated by an anticorrelation with \cfe , as the
index is also expected to increase with carbon abundance (see pink and
green solid curves in Fig.~\ref{fig:atio_nocorr}). Notice that since
the radial trends of \afe\ and \cfe\ are rather flat in XSG1, based on
CvD12 model predictions, one would not expect any gradient
of \atio\ throughout the galaxy.  The strong radial variation of
\atio\ in Fig.~\ref{fig:atio_nocorr} seems thus unexplained. This is
further shown in Fig.~\ref{fig:atio_corr}, where empty circles plot
line-strengths corrected to the solar scale with CvD12 models.  One can
explain the innermost corrected value with MIUSCAT models 
but there is no model prediction matching the outermost data-points.
However, one should consider that (i) CvD12 predictions for varying
abundance ratios are only available for SSPs with solar
metallicity, and a Chabrier IMF. In practice, the effect of
\afe\ and \cfe\ might be dependent on metallicity and IMF; (ii)
\atio\ is a broad feature, and its sensitivity to \cfe\ may be more
uncertain, {\it as it is entirely due to a variation of the continuum in
the CvD12 models } (see green dashed line in Fig.~\ref{fig:atio_corr});
(iii) in our data, the absolute value of \atio\ is affected by the
time-varying response of the UVB--VIS X-SHOOTER dichroic. We measured
this effect by computing the RMS of \atio\ values, in the central
spectrum of XSG1, among different exposures (see grey double-sided
arrow at top--right of Fig.~\ref{fig:atio_corr}).  To gain insight
into these issues, we also plot in Fig.~\ref{fig:alpha} the
empirically abundance-corrected values of \atio\ (see empty squares in
the Figure).
 Indeed,
following the approach of LB13, we find that, independently of the
bin in velocity dispersion from the SDSS spectra, \atio\ increases with \afe , as
$\rm \delta_{aTiO}=\delta(aTiO)/\delta(\alpha/Fe) \sim
0.006~mag/dex$. This implies a shift of $\sim -0.0025$~mag (for
\afe$\sim 0.4$) for all, but the outermost, uncorrected values of
\atio\ (Fig.~\ref{fig:atio_nocorr}). However, the value of
\datio\ might be underestimated for the innermost data-points of
XSG1. At fixed velocity dispersion in the SDSS data, there is also a
mild increase ($\sim 0.1$~dex) of metallicity with \afe\ -- an effect
that we have removed following LB13, by subtracting off the index variation
expected from the change in metallicity for Kroupa-like MIUSCAT models
(see sec.~5 of LB13). While this is correct for indices whose
sensitivity to \zh\ does not depend on IMF (e.g. \tioi\ and \tioii ),
it is not for \atio . Accounting for this effect, would make our
empirical correction larger, by a factor of $\sim 3$, in the innermost
point of XSG1, implying the corrected value of \atio\ in the centre
to be $\sim 0.01$~mag, in very good agreement with results from our
best-fitting procedure to applied all optical indices
(Sec.~\ref{sec:gamma_trends}; Fig.~\ref{fig:ind_rad}). In summary, we
think that the large radial gradient of \atio\ in our target galaxy is
a combination of (1) an underestimation of our empirical correction at
high metallicity (we are currently working to improve method in this
regard), and (2) the uncertainty in the zero-point of the index due to
the dichroic response. Regarding point (1), it does not affect
our results (see Sec.~\ref{sec:gamma_trends}), whereas for point (2), we
are currently working to improve the treatment of the dichroic issue
in the reduction procedure. For the present work, we have decided
to simply add in quadrature the uncertainty on the dichroic response to
the observed error bars on \atio .

\subsection{The Mg4780 index}

\begin{figure}
\begin{center}
\leavevmode
\includegraphics[width=8.5cm]{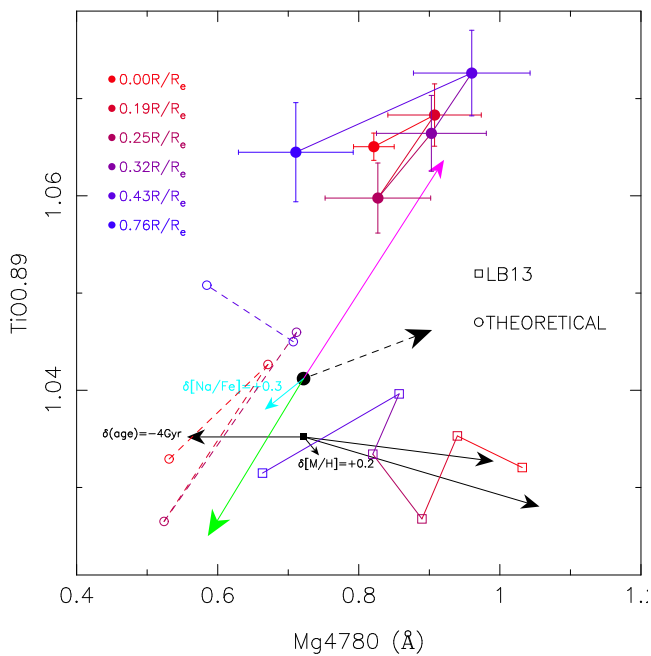}
\end{center}
\caption{Same as in Fig.~\ref{fig:tio_nocorr} but plotting \tionir\ as
  a function of the \mgf\ IMF-sensitive feature. The cyan arrow shows
  the sensitivity of \mgf\ on [Na/Fe], based on CvD12 SSP models.  }
\label{fig:mgf_nocorr}
\end{figure}

Fig.~\ref{fig:mgf_nocorr} plots the \tionir\ as a function of \mgf
. Interestingly, no measurable radial gradient is found in \mgf . The
index tends to increase radially, from the innermost to the second
outermost bin, while the outermost bin value is fully consistent with
the central one (see blue and red solid circles in the Figure).  Using
CvD12 models, we find that the line-strength of \mgf\ is expected to
increase with \afe\ (mostly because of the sensitivity to \mgfe ,
consistent with~\citealt{Serven:2005}) and decrease with \cfe ,
similarly to \tioi\ and \tioii . Subtracting off the expected
variation of the index from our \afe\ and \cfe\ estimates
(Fig.~\ref{fig:abund}) with CvD12 models, we obtain the
``theoretically-corrected'' values (empty circles) in
Fig.~\ref{fig:mgf_nocorr}, still showing (as expected) no significant
radial gradient. Notice that the theoretically-corrected
line-strengths are only consistent with a Kroupa-like IMF (especially
in the innermost bin) and an age of $\sim 10$--$11$\,Gyr, inconsistent
with results for \tioi\ and \tioii .  MN15a already noticed
that \mgf\ is also dependent (to a lesser extent than \afe\ and \cfe )
on other single-element abundance ratios, i.e. \nafe\ and [Si/Fe]. In
particular, \mgf\ is expected to be {\it anti-correlated } with
\nafe\ because of its sensitivity to sodium on the blue
pseudo-continuum, as illustrated by the cyan arrow in
Fig.~\ref{fig:mgf_nocorr}. Hence, a large \nafe\ radial-gradient would
tend to cancel out the effect of an IMF-gradient on \mgf . The empty
squares in Fig.~\ref{fig:mgf_nocorr} illustrate the results of our
empirical correction procedure. As discussed in LB13, quite
unexpectedly, the index tends to increase with \afe\ in the SDSS
stacks at the lowest velocity dispersion (metallicity), while it {\it
  decreases} with \afe\ at the other end. As a result, the range of
values for \mgf\ as a function of $\sigma$ (see panel b in fig.~8 of
LB13), is significantly larger at high \afe , than at \afe$ \sim
0$. This is fully consistent with the fact that, at such high
\afe\ ($\sim 0.4$), as is the case in XSG1, we do not see any radial gradient in
\mgf . Furthermore, in the massive ETG analyzed in MN15a we saw
\mgf\ to feature a shallower radial gradient than other IMF indicators
(mostly \tioii ).  Remarkably, our empirical correction -- performed
as a function of metallicity -- recovers a clear radial gradient
in XSG1, with values fully consistent with those obtained from the
other optical indicators (see Sec.~\ref{sec:gamma_trends}. Perhaps,
the inconsistency between theoretical and empirical corrections for
\mgf\ arise from the fact that some other elements (e.g. sodium) 
also give a prominent contribution to the observed line-strengths,
and/or the fact that the effect of \afe\ and \cfe\ is strongly
dependent, for this index, on \zh -- something that cannot be tested
with the available CvD12 models.
  
\label{lastpage}

\end{document}